\documentclass{aa}
\bibpunct{(}{)}{;}{a}{}{,}
\usepackage[varg]{txfonts}
\usepackage{graphicx,gensymb}
\usepackage{subcaption}
\usepackage[normalem]{ulem}
\usepackage{amsmath, amssymb, amsfonts}
\usepackage{rotating} 
\usepackage{adjustbox}
\usepackage{ulem}
\begin{document}

\title{The nature of hyperluminous infrared galaxies}

\author{F.~Gao\inst{\ref{inst1},\ref{inst2}}
\and L.~Wang\inst{\ref{inst1},\ref{inst2}}
\and A.~Efstathiou \inst{\ref{inst3}}
\and K.~Ma\l ek \inst{\ref{inst4},\ref{inst5}}
\and P.~N.~Best\inst{\ref{inst6}}
\and M.~Bonato\inst{\ref{inst7},\ref{inst8},\ref{inst9}}
\and D.~Farrah\inst{\ref{inst10},\ref{inst11}}
\and R.~Kondapally\inst{\ref{inst6}}
\and I.~McCheyne\inst{\ref{inst12}}
\and H. J. A.~R$\rm \ddot{o}$ttgering\inst{\ref{inst13}}
}

\institute{Kapteyn Astronomical Institute, University of Groningen, Postbus 800, 9700 AV Groningen, The Netherlands\label{inst1}
\and 
SRON Netherlands Institute for Space Research, Landleven 12, 9747 AD, Groningen, The Netherlands\label{inst2}
\and
School of Sciences, European University Cyprus, 6, Diogenes Street, Engomi, 1516 Nicosia, Cyprus\label{inst3}
\and
National Centre for Nuclear Research, ul. Pasteura 7, 02-093 Warszawa, Poland\label{inst4}
\and
Aix Marseille Univ. CNRS, CNES, LAM, Marseille, France\label{inst5}
\and 
Institute for Astronomy, University of Edinburgh, Royal Observatory, Blackford Hill, Edinburgh, EH9 3HJ, UK\label{inst6}
\and
Italian ALMA Regional Centre, Via Gobetti 101, I-40129, Bologna, Italy\label{inst7}
\and
INAF-IRA, Via Gobetti 101, I-40129, Bologna, Italy\label{inst8}
\and
INAF-Osservatorio Astronomico di Padova, Vicolo dell'Osservatorio 5, I-35122, Padova, Italy\label{inst9}
\and
Department of Physics and Astronomy, University of Hawaii, 2505 Correa Road, Honolulu, HI 96822, USA\label{inst10}
\and
Institute for Astronomy, 2680 Woodlawn Drive, University of Hawaii, Honolulu, HI 96822, USA\label{inst11}
\and
Astronomy Centre, Department of Physics \& Astronomy, University of Sussex, Brighton, BN1 9QH, England\label{inst12}
\and
Leiden Observatory, Leiden University, PO Box 9513, 2300 RA Leiden, the Netherlands\label{inst13}
}

\abstract{Hyperluminous infrared galaxies (HLIRGs) are shown to have been more abundant in early epochs. The small samples used in earlier studies are not sufficient to draw robust statistical conclusions regarding the physical properties and the power sources of these extreme infrared (IR) bright galaxies.}
{We make use of multi-wavelength data of a large HLIRG sample to derive the main physical properties, such as stellar mass, star formation rate (SFR), volume density, and the contribution to the cosmic stellar mass density and the cosmic SFR density. We also study the black hole (BH) growth rate and its relationship with the SFR of the host galaxy.}
{We selected 526 HLIRGs in three deep fields (Bo$\rm \ddot{o}$tes, Lockman-Hole, and ELAIS-N1) and adopted two spectral energy distribution (SED) fitting codes: CIGALE, which assumes energy balance, and CYGNUS, which is based on radiative transfer models and does not adopt an energy balance principle. We used two different active galactic nucleus (AGN) models in CIGALE and three AGN models in CYGNUS to compare results that were estimated using different SED fitting codes and a range of AGN models.}
{The stellar mass, total IR luminosity, and AGN luminosity agree well among different models, with a typical median offset of 0.1 dex. The SFR estimates show the largest dispersions (up to $0.5$ dex). This dispersion has an impact on the subsequent analysis, which may suggest that the previous contradictory results could partly have been due to the different choices in methods. HLIRGs are ultra-massive galaxies, with $99\%$ of them having stellar masses larger than $10^{11} M_{\odot}$. Our results reveal a higher space density of ultra-massive galaxies than what was found by previous surveys or predicted via simulations. We find that HLIRGs contribute more to the cosmic SFR density as redshift increases. In terms of BH growth, the two SED fitting methods provide different results. We can see a clear trend in whereby SFR decreases as AGN luminosity increases when using CYGNUS estimates. This may possibly imply quenching by AGN in this case, whereas this trend is much weaker when using CIGALE estimates. This difference is also influenced by the dispersion between SFR estimates obtained by the two codes.}{}

\maketitle
\thispagestyle{empty}
\section{Introduction}

Around half of the emitted energy coming from galaxies in the Universe is absorbed by dust and then re-emitted at infrared (IR) wavelengths \citep{1998ApJ...508..106D, 1998ApJ...508..123F, 2010PASP..122..499E, 2012MNRAS.424.1614O,2014PhR...541...45C}. Since the first all-sky survey carried out by Infrared Astronomical Satellite (IRAS), we have uncovered many IR galaxies that are faint in optical surveys \citep{1984ApJ...278L...1N, 1985ApJ...290L...5H, 1987ApJ...320..238S,1996ARA&A..34..749S}, emitting the bulk of their energy in the IR from dust heated by young stars in dusty star-forming regions, and/or from dust in a torus-like structure surrounding a central supermassive black hole in active galactic nuclei (AGN). These IR luminous galaxies are important for studying the cosmic star-formation history and the co-evolution of black holes and host galaxies because they play a key role in measuring star formation and nuclear activity obscured by dust and are abundant at high redshifts\citep{1998Natur.394..241H, 1998Natur.394..248B, 2005ApJ...631L..13D, 2013MNRAS.432...23G, 2014ARA&A..52..415M}. 

Such IR luminous galaxies can be separated into different populations according to their IR luminosity, $L_{IR}$, {integrated from rest-frame 8-1000 $\mu m$}: luminous IR galaxies (LIRGs) with $L_{IR}>10^{11} L_{\odot}$, ultraluminous IR galaxies (ULIRGs) with $L_{IR}>10^{12} L_{\odot}$, and hyperluminous IR galaxies (HLIRGs) with $L_{IR}>10^{13} L_{\odot}$ \citep{2000MNRAS.316..885R}. While these IR luminous galaxies are relatively rare in the local universe, studies at higher redshifts show a prominent population of LIRGs at z$\sim 1$ and ULIRGs at z$\sim 2$ \citep{2001ApJ...556..562C, 2005ApJ...632..169L, 2013A&A...553A.132M, 2015A&A...575A..74S}. Previous studies have mostly focused on LIRGs and ULIRGs, finding that LIRGs at z $\sim$ 1 and ULIRGs at z $\sim$ 2 show similar properties to local normal star-forming galaxies, while only the starbursts lying significantly above the SFR-M$_*$ main sequence (MS) correlation \citep{2004MNRAS.351.1151B, 2007ApJ...660L..43N, 2007A&A...468...33E} can be regarded as true analogs of local IR luminous galaxies \citep{2015A&A...575A..74S}. However, such HLIRGs are extremely rare and IR luminous, which means that we need to probe large volumes to find a significant number of these galaxies.

IRAS 10214+4724 was one of the first HLIRGs detected by IRAS \citep{1991Natur.351..719R, 1992ApJ...399L..55S}, lying at $z= 2.286$. Submillimeter observations revealed a significant detection, suggesting that this source is undergoing extreme star-burst activity \citep{1992MNRAS.256P..35C}, possibly supported by its large molecular gas content \citep{1992ApJ...398L..29S}. Optical and near-IR images have shown that this galaxy is gravitationally lensed and has magnified brightness \citep{1995ApJ...450L..41B,1995ApJ...449L..29G}. However, even after correcting for  amplification, it still exhibits an intrinsic IR luminosity of $2 \times 10^{13} L_{\odot}$ \citep{1996ApJ...461...72E}. Highly polarized continuum and broad emission lines suggest that IRAS 10214+4724 could contain a hidden quasar \citep{1995AAS...18711902G}. 

Submillimeter observations play an important role in detecting high redshift IR luminous galaxies because the rest-frame FIR tail shifts to the longer wavelength submillimeter range. SMM J02399-0136 at $z\sim 2.8$ was the first submillimeter-detected HLIRG, showing broad emission line widths as well as line ratios indicative of a dust-reddened AGN \citep{1998MNRAS.298..583I}. Follow-up multi-wavelength data revealed a massive, obscured starburst that dominates the FIR emission \citep{2010MNRAS.404..198I}.


The relative contribution of starburst and AGN to the luminosity of IR luminous galaxies is the subject of ongoing debate. Early studies found an increase of AGN fraction as the total energy output increases, based on AGN selections using optical classification \citep{1999ApJ...522..113V} and mid-IR spectroscopy \citep{2001ApJ...552..527T}. 

\citet{2002MNRAS.335.1163F} found that both the AGN and starburst component are important when fitting the IR emission of a sample of 11 HLIRGs, with the AGN component varying in importance from 20\% to 80\%. \citet{2002MNRAS.335..574V} reached a similar conclusion by studying five HLIRGs detected by the  \textit{Infrared Space Observatory}. 
\citet{2013A&A...549A.125R} found that AGN are dominant based on MIR analysis using a sample of 13 HLIRGs. These studies also showed that the starburst contribution is significant, with a mean contribution of 30\%, suggesting that both AGN and starburst are contributing to the IR brightness, but their relative dominance may vary a lot on an object by object basis.

A significant number of massive star-forming and quiescent galaxies found at z$>3$ \citep{2014ApJ...787L..36S, 2018A&A...618A..85S, 2019Natur.572..211W}  may raise tension with the hierarchical clustering scenario, in which small halos form earlier and coalesce later to form large halos \citep[e.g.,][]{1991ApJ...379...52W, 1993MNRAS.262..627L, 1997ApJ...490..493N,1999MNRAS.303..188K, 2000MNRAS.319..168C}. Current galaxy formation models have not been especially successful in producing a sufficient number of high-redshift massive galaxies. Massive star-forming $z\sim2$ galaxies are found to be common in ULIRG mode \citep{2007ApJ...670..173D, 2007ApJ...670..156D}. Therefore, HLIRGs, being both extremely IR luminous and massive, are expected to play an important role in understanding the evolution of high redshift ultra-massive star-forming galaxies.


In this paper, we present the largest sample of HLIRGs, containing 526 sources in the Bo$\rm \ddot{o}$tes, Lockman-Hole (LH), and ELAIS-N1 (EN1) fields. We study their properties and their nature in a cosmological context. We briefly introduce our sample selection in Section \ref{data} and describe our spectral energy distribution (SED) fitting methods in Section \ref{method}. Our results are presented in Section \ref{re}, where we first compare estimates derived from different SED fitting codes and AGN models. Then we study the stellar mass properties of our HLIRG sample and their co-moving volume density, followed by analyses of their star-formation activity. We also study the AGN activity of our HLIRG sample and the connection between AGN activity and star-formation activity. Finally, we summarize our conclusions in Section \ref{conclu}. Throughout this paper, we assume a flat $\Lambda$CDM universe with $\Omega_{\text{M}} = 0.286$ and $H_{0}=69.3 \,\rm km s^{-1} Mpc^{-1}$. Unless otherwise stated, we adopt a \citet{1955ApJ...121..161S} initial mass function (IMF).

\section{Data}\label{data}
In this section, we briefly describe  how we built our HLIRG sample. We note that the full details of the construction of our parent sample are presented in \citet{2021A&A...648A...8W}. In summary, our parent sample is based on a \textit{Herschel} selection at 250 $\mu m,$ which was then cross-matched to the radio 150 MHz imaging data from the Low Frequency Array (LOFAR), specifically, the multi-wavelength photometry and redshift information from the LOFAR Two-metre Sky Survey (LoTSS) Deep Fields First data release \citep{2021A&A...648A...4D,2021A&A...648A...3K, 2021A&A...648A...2S, 2021A&A...648A...1T}, in order to find the right multi-wavelength counterparts of the \textit{Herschel} sources. \citet{2021A&A...648A...3K} provided robust cross-identifications and new forced, matched-aperture multi-wavelength photometry for radio detected sources in three LOFAR deep fields. These authors successfully identified multi-wavelength counterparts for $>97\%$ radio detected sources, using a combination of Likelihood Ratio method \citep{1977A&AS...28..211D, 1992MNRAS.259..413S} and visual classification. \citet{2021A&A...648A...4D} presented consistent photometric redshift estimates for sources in three deep fields, using a hybrid methodology that combines template fitting and machine learning. These photometric redshift estimates show a 1.6-2\% and 6.4-7\% scatter from spectroscopic redshifts for galaxies and AGNs, respectively. The outlier fractions (defined as |$\Delta$z|/(1+spec-z)>0.15) are 1.5-1.8\% and 18-22\% for galaxies and AGNs, respectively.

In order to find multi-wavelength counterpart for each \textit{Herschel} source, we first cross-matched the \textit{Herschel}-SPIRE blind catalog to the LoTSS Deep Fields catalog and adopted a flux density cut to select SPIRE blind sources, requiring a 250 $\mu m$ flux density higher than 35, 40, 45 mJy in EN1, LH, and Bo$\rm \ddot{o}$tes fields, respectively. Then we separated SPIRE blind sources into a unique sample and a multiple sample, which corresponds to single and multiple LOFAR counterparts within 18\arcsec matching radius (Herschel 250 $\mu m$ beam size), respectively. For the unique sample, we can directly apply the \textit{Herschel} fluxes from the blind catalog. For the multiple sample, we need to de-blend the \textit{Herschel} fluxes for each component corresponding to the same \textit{Herschel} source. By exploiting the FIR-radio correlation (i.e., the tight relationship between LOFAR 150 MHz and \textit{Herschel} flux densities, e.g., \citet{2018MNRAS.475.3010G, 2019A&A...631A.109W, 2021A&A...648A...6S}), we can estimate the expected \textit{Herschel} fluxes for the multiple sample. These expected fluxes serve as a prior in the de-blending process using XID+ software \citep{2017MNRAS.464..885H},  which is a probabilistic de-blending tool to calculate the de-blended \textit{Herschel} fluxes for the multiple sample. 

After running a preliminary SED fitting procedure, we selected 201, 269, and 72 HLIRGs that have total IR luminosity $L_{IR}>10^{12.95} L_{\odot}$ (i.e., a slightly lower threshold than typical HLIRGs) in the EN1, LH, and Bo$\rm \ddot{o}$tes fields, respectively. We further excluded three, ten, and three HLIRGs that have photometric redshifts >6, considering the phot-z quality at such high redshifts \citep[see][]{2021A&A...648A...4D}; this resulted in 198, 259, and 69 HLIRGs  in the EN1, LH, and Bo$\rm \ddot{o}$tes fields respectively. \citet{2021A&A...648A...8W} selected HLIRGs using an IR luminosity that is strictly resulting from star formation. In this paper, we also include contribution from AGN and we slightly lower the threshold of IR luminosity, resulting in 73, 95, 25 more HLIRGs in the EN1, LH, and Bo$\rm \ddot{o}$tes fields, respectively. 

With this paper, we publish our sample of HLIRGs, together with the derived physical properties. We list examples of our HLIRGs in Table \ref{source_list} in the appendix. The full list of HLIRGs will be given online. The data products include sky positions, redshift (spectroscopic or photometric), and galaxy properties such as the stellar mass, SFR, total IR luminosity, and AGN fraction derived using different methods (see next section).

\section{Methods}\label{method}
We employed two sets of SED fitting methods: 
Code Investigating GALaxy Emission (CIGALE) and CYprus Models for Galaxies and their NUclear Spectral (CYGNUS). CIGALE is a widely adopted SED fitting codes that is based on the assumption of energy balance. Energy balance means that the energy absorbed by dust in the ultraviolet (UV)-optical-near IR band is re-emitted self-consistently in the mid- and far-IR range \citep{Boquien19}. There are other SED fitting methods that utilize the same principle, such as MAGPHYS \citep{Dacunha}, Le PhARE \citep{2011ascl.soft08009A}, and Prospector \citep{2017ApJ...837..170L}. We chose CIGALE over MAGPHYS as the latter  does not include AGN models that are proven to be non-negligible when fitting HLIRGs \citep[e.g.,][]{2002MNRAS.335.1163F, 2017ApJ...844..106F, 2018MNRAS.479L..91S}. However, some recent studies have shown that the energy balance principle may not properly account for the IR luminosity of very dusty IR luminous galaxies because dust and stellar distributions in these galaxies may not be well connected. For example, \citet{2019A&A...632A..79B} studied a sample of 17 Atacama Large Millimeter Array (ALMA) and Herschel detected sources at $z\sim2$, finding that when fitting the stellar continuum with attenuation laws, only one-third to half of the total dust emission detected by ALMA and \textit{Herschel} can be recovered. Considering this potential issue, we also took advantage of CYGNUS, which is based on radiative transfer models and does not adopt the energy balance principle. Another advantage is that CYGNUS can account for a finer grid of parameters of the AGN model, while CIGALE only provides limited options in the AGN module. The AGN models in CYGNUS in this method have been proven to fit extreme IR luminous galaxies successfully in both local \citep[][Efstathiou et al. in preparation]{2003MNRAS.343..585F, 2014MNRAS.437L..16E} and high-redshift systems \citep{2017ApJ...844..106F, E13, 2021MNRAS.503L..11E}.

\subsection{The CIGALE code}
CIGALE was first written by \citet{Burgarella} and then further developed by \citet{Noll}. The fitting process is as follows: stellar population models are built first and then reddened by dust. The absorbed energy is re-emitted in the IR band, in which dust emission due to AGN is added. The best-fit model is selected by directly comparing model fluxes and input observational data according to the goodness of fit. The galaxy properties are not estimated simply from the best-fit model but based on a Bayesian-like approach to calculate the probability distributions \citep{Noll, Boquien19}. CIGALE has been applied successfully in many studies, such as dust attenuation \citep{Buat, Salim}, the properties of AGN host galaxies \citep{Ciesla}, the IRX-$\beta$ relationship in star-forming galaxies \citep{beta}, star-forming main sequence \citep{2015MNRAS.453.2540J, Pearson18}, SFR estimators \citep{2014A&A...561A..39B, 2016A&A...591A...6B}, and the global characteristic of millions of IR detected galaxies in the Herschel Extragalactic Legacy Project 
\citep[HELP][]{2018A&A...620A..50M}. Recently, the X-CIGALE version was released including the X-ray domain to fit multi-wavelength data from X-ray to FIR \citep{2020MNRAS.491..740Y}, however, we did not use this version of CIGALE in this study.

We used a delayed-$\tau$ (almost linearly increase in SFR with time, followed by a smooth decrease after $t=\tau$) plus a starburst star formation history (SFH) and \citet{bc03} single stellar populations (SSPs), with a Salpeter initial mass function (IMF) and solar metallicity. The delayed- $\tau$ model is successful in modeling both early-type (with small $\tau$) and late-type (with large $\tau$) galaxies \citep{Boquien19}. We added a recent starburst component to model a young star-forming population. We adopted a double power-law dust attenuation law based on \citet{CF00}, one power-law for the birth cloud (BC) and the other for the interstellar medium (ISM). With this recipe, young stars are attenuated twice: by the surrounding BC dust and additionally by the dust in the ISM. In terms of dust emission, we utilized the models present in \citet{DL14} as they take into account a wide range of radiation fields and polycyclic aromatic hydrocarbon (PAH) emission \citep{Boquien19}.

In order to investigate the impact of different AGN templates, we use two sets of AGN templates provided in CIGALE, both based on theoretical models but assuming different distributions for the dust in the torus. One is the \citet[][denoted as the \textbf{F06} model thereafter]{Fritz} smooth AGN templates assuming a flared disk (having a simple structure with polar opening) torus structure and a wide range of grain sizes of graphite and silicate. The SEDs are derived by calculating 2-D radiative transfer equations. The other set is the SKIRTOR clumpy AGN templates proposed by \citet[][denoted as the \textbf{S12} model thereafter]{Stalevski}, which consider a two-phase model with high-density clumps and low-density medium filling between clumps in their 3-D radiative transfer modeling. The smooth model requires that the dust density can only vary continuously within the torus while the clumpy model suggests that dust is distributed in clumps. \citet{2002ApJ...570L...9N, 2008ApJ...685..147N, 2008ApJ...685..160N} claimed that a 3D radiative transfer of a clumpy torus model can better reproduce the observed properties of the silicate feature than that of a smooth model, while 2D radiative transfer modeling provided by \citet{2005A&A...436...47D} disagrees with that claim. \citet{2012MNRAS.426..120F} performed a thorough comparison between the smooth models of \citet{Fritz} and the clumpy models of \citet{2002ApJ...570L...9N, 2008ApJ...685..147N, 2008ApJ...685..160N}, finding distinct SEDs even with matched parameters. However, these differences are due to model assumptions rather than to the issue of whether the dust distribution is clumpy or smooth. In addition, ambiguities make it hard to discriminate between these two models. In our work, we provide CIGALE fitting results using both models. The parameter space in the two CIGALE runs is listed in Table \ref{para_CIGALE} in the appendix.

\subsection{The CYGNUS code}
CYGNUS is a collection of libraries of radiative transfer models combining AGN tapered disks and tori models \citep[the height of the disk increases as the distance from the center increases but remains fixed to a constant value in the outer part;][denoted as the \textbf{E95} model thereafter]{ER95, E13}, starburst galaxies \citep{E00, E09}, and host galaxies \citep{ER03, 2021MNRAS.503L..11E}. We also include another two sets of AGN models in the SED fitting with CYGNUS: the above-mentioned F06 and S12 AGN models.

These individual models have been used successfully in various studies of, for instance, star-forming galaxies \citep{1997MNRAS.289..490R}, Seyfert galaxies \citep{2003AJ....126...81A}, local LIRGs \citep{2017MNRAS.471.1634H} and ULIRGs \citep{2003MNRAS.343..585F}, high-z submillimeter detected galaxies \citep{ER03}, radiatively driven outflows in QSOs \citep{2012ApJ...745..178F}, high-z HLIRGs \citep{2002MNRAS.335.1163F, 2002MNRAS.335..574V, 2017ApJ...844..106F}, and so on. The starburst models are a combination of the evolution of giant molecular clouds as HII regions expand due to ionization, \citet{bc03} SSPs, and a detailed radiative transfer that accounts for the effect of PAH and small grains. 
CYGNUS models the starburst component with optically thick radiative transfer models. The models of \citet{DL14} used by CIGALE are not computed with a radiative transfer code but assume optically thin emission from the dust as they were designed to model the emission of the Andromeda galaxy.

The host galaxy models are dominated by emission from dust with low optical depth and cool (<30 K) temperatures \citep{ER03}. In this work, we model the host galaxy using libraries of spheroidal models computed at different redshifts as described in \citet{2021MNRAS.503L..11E}.

\begin{figure}
\resizebox{\hsize}{!}{\includegraphics[width=\linewidth]{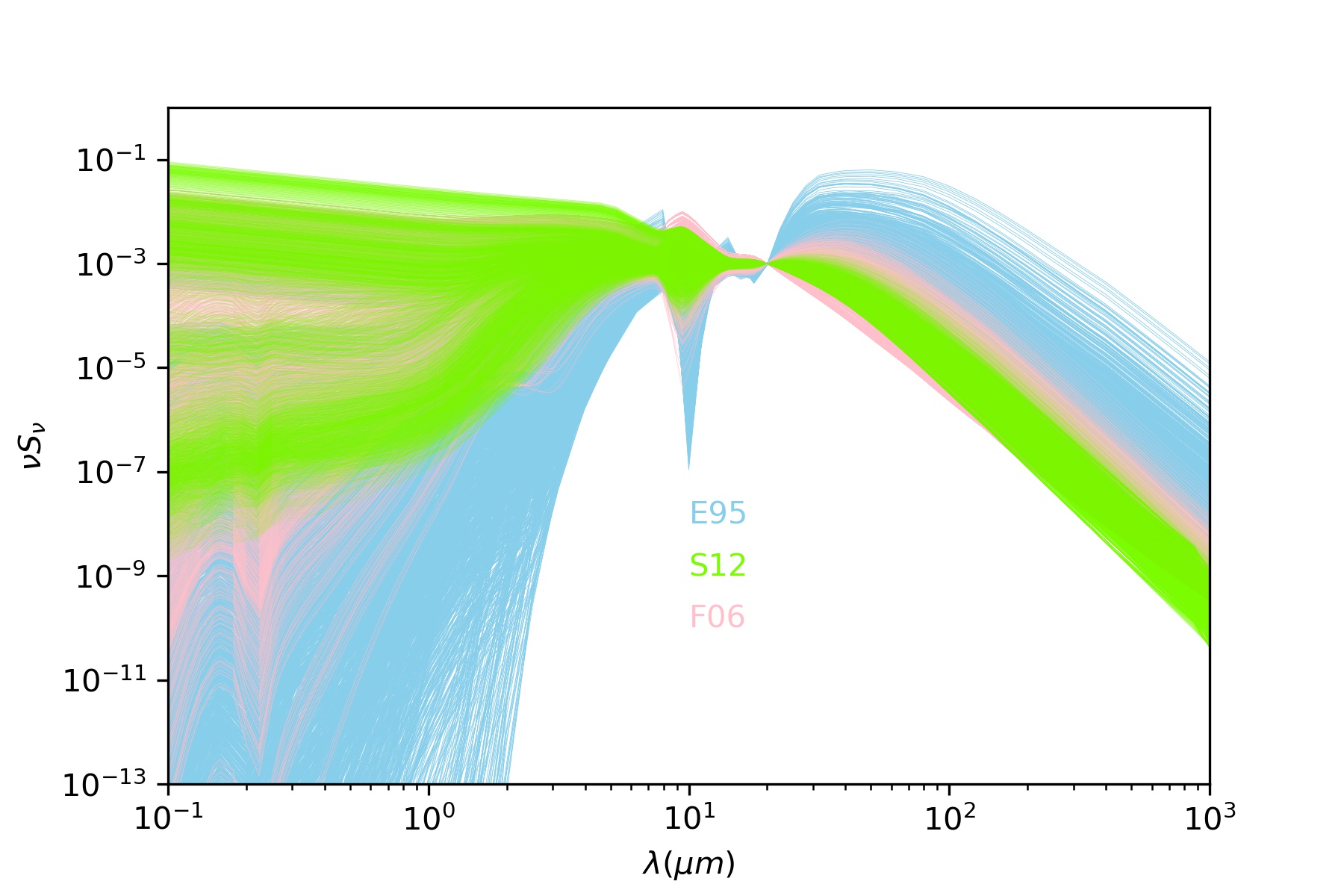}}
\caption{Three AGN libraries \citep{ER95, Fritz, Stalevski} used in our work, normalized at 20 $\mu m$.}
\label{agn_temp}
\end{figure}

In Figure \ref{agn_temp}, we show three AGN models that we use in two SED fitting methods, normalized at 20 $\mu m$. The E95 model encompasses a wider range of spectral shapes than the other two AGN models in the longer wavelength range. Figure \ref{example} in the appendix illustrates examples of best-fit SEDs using CIGALE F06 and CYGNUS F06 models for the same galaxy. We show example galaxies for which both SED fitting codes can fit well ($\sim84\%$ in the total sample); as well as where the CIGALE model fails to provide good fits and CYGNUS can ($\sim8\%$ in the total sample);  or neither provide a good fit ($\sim6\%$ in the total sample).

\begin{figure*}
\resizebox{\hsize}{!}{\includegraphics[width=\linewidth]{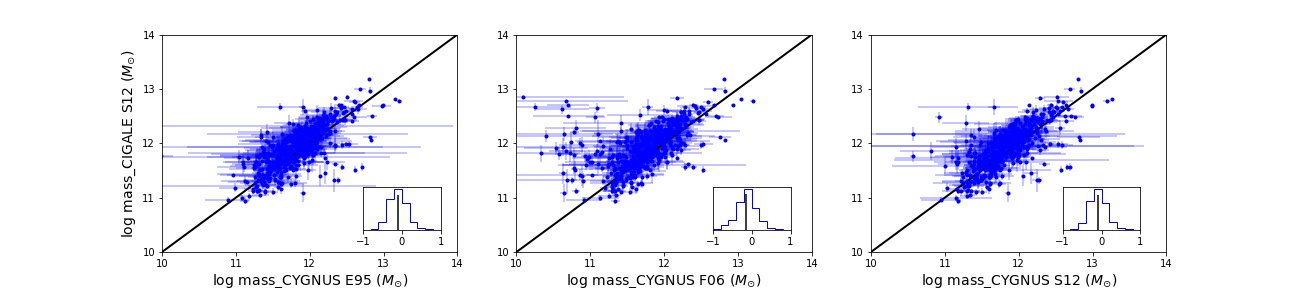}}
\caption{Comparisons of stellar mass estimates derived using the CIGALE S12 model and the three AGN models in the CYGNUS runs (in the order of E95, F06, and S12). The histograms corresponding to the difference between the X- and Y-axis values are inserted in each panel.}
\label{compare_mass}
\end{figure*}

\begin{figure*}
\resizebox{\hsize}{!}{\includegraphics[width=\linewidth]{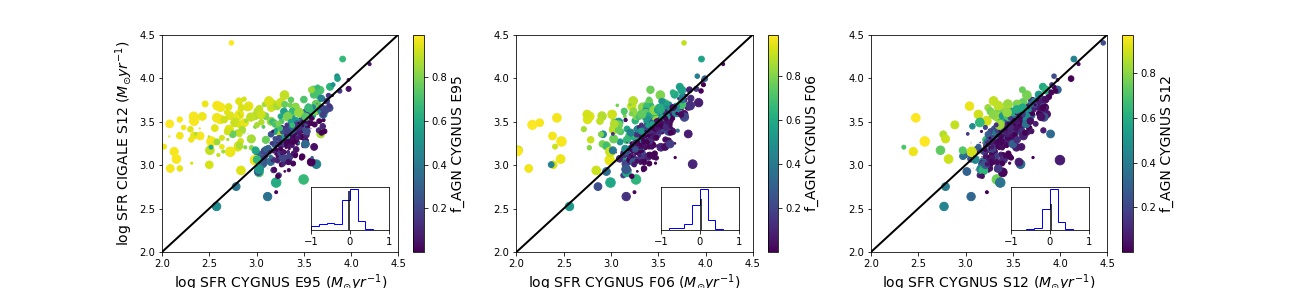}}
\caption{Comparisons of SFR estimates derived using the total IR luminosity, excluding the AGN contribution between the CIGALE S12 and the three AGN models in the CYGNUS runs. The colors denote AGN fractions derived using the three AGN models in CYGNUS (left: E95; middle: F06; right: S12) and the sizes denote AGN fractions derived using CIGALE S12. The histograms corresponding to the difference between the $x$- and $y$-axis values are inserted in each panel.}
\label{compare_sfr}
\end{figure*}

\begin{figure*}
\resizebox{\hsize}{!}{\includegraphics[width=\linewidth]{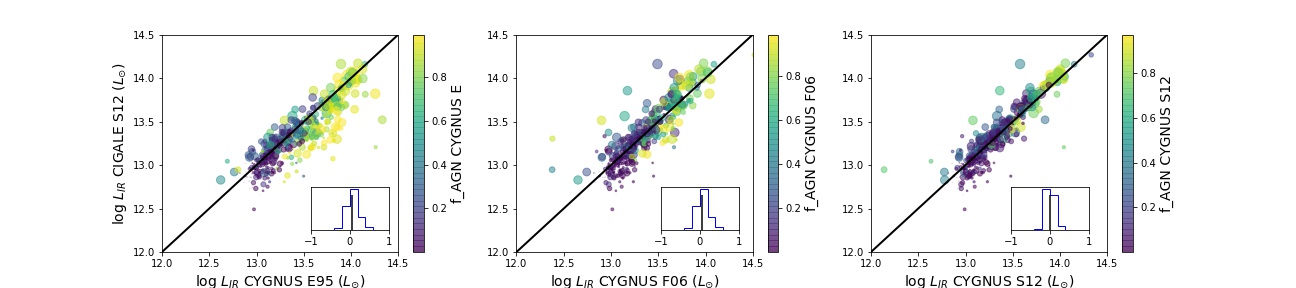}}
\caption{ Comparisons of total IR luminosity estimates derived using the CIGALE S12 and the three AGN models in the CYGNUS runs. }
\label{compare_lir}
\end{figure*}

\begin{figure*}
\resizebox{\hsize}{!}{\includegraphics[width=\linewidth]{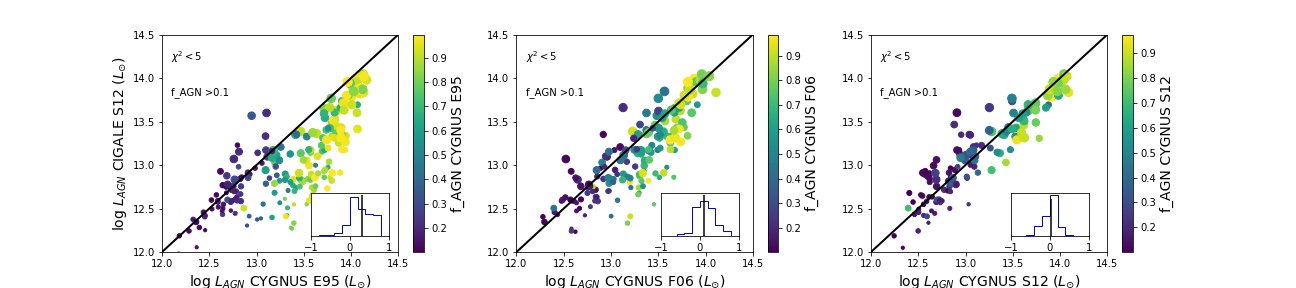}}
\caption{Comparisons of AGN luminosity estimates derived using the CIGALE S12 and the three AGN models in the CYGNUS runs. Only sources having AGN fraction $>0.1$ and reduced $\chi^2<5$ for both methods participate in this comparison.}
\label{compare_lagn}
\end{figure*}

\section{Results}\label{re}
\subsection{Comparisons of galaxy properties from different SED fitting tools and AGN models}\label{comp}
In this section, we compare the galaxy properties derived from different SED fitting methods and AGN models. In Figures \ref{compare_mass}, \ref{compare_sfr}, \ref{compare_lir}, and \ref{compare_lagn}, we only provide comparisons between the CIGALE run using the S12 model and CYGNUS runs using three AGN models; this is due to the similarities among the results obtained in two CIGALE runs. In general, little to no systematic offset exists in all five fitting estimates between CIGALE and CYGNUS runs, although they can span a broad range around the one-to-one line. 

The stellar mass estimates agree well among different models, with the maximum median difference $\sim$0.1 dex. Differences arise due, in turn, to the differences in the SED fitting codes as well as differences in the adopted AGN models. There are some sources that have large error bars estimated in the CYGNUS runs. This is because CIGALE uses a finer grid of parameters in the stellar populations, while CYGNUS uses denser parameters in AGN templates, resulting in some galaxies having stellar masses that are not well constrained in CYGNUS. 

In the SED fitting codes CIGALE and CYGNUS, there are several SFR indicators based on different timescales. For example, CIGALE provides instantaneous SFR, as well as SFRs that are averaged over the last 10 Mega years and 100 Mega years. It is widely accepted to use SFR averaged over the last 100 Mega years (hereafter denoted as SFR$_{100 Myr}$) to represent a stable star-formation activity. CYGNUS also provides a SFR that is averaged over burst age of the younger stellar population (hereafter denoted as SFR$_{burst\,age}$).We compare these two SFRs (i.e., SFR$_{100 Myr}$ and SFR$_{burst\,age}$) with IR luminosity that is only due to star formation in Figure \ref{timescale_comp} in the appendix, finding that SFR$_{burst\,age}$ shows a smaller scatter. This is expected as burst age is usually less than 100 mega years. Using different SFRs may influence some of our results. We briefly discuss this influence in Appendix \ref{timescale}. In our main text, to obtain a homogeneous parameter when making comparisons between both codes, we adopt SFRs based on IR luminosity that is only due to star-formation (i.e., excluding AGN contribution) using the law of \citet{1998ARA&A..36..189K}:

\begin{equation}
SFR (M_{\odot}\, yr^{-1}) = 4.5 \times 10^{-44} L_{IR, SF} (\rm ergs\, s^{-1}). 
\end{equation}
The SFRs do not show significant median offset but they do have a large 1-$\sigma$ dispersion between CIGALE and CYGNUS runs. In general, sources for which CYGNUS predicts a high AGN fraction tend to have lower SFRs compared to CIGALE while sources with low AGN fractions in CYGNUS have relatively higher SFRs compared to CIGALE. Within the CYGNUS runs with three AGN models, the E95 model shows a longest tail toward low SFR end. We visually inspect the best-fit plots from all CYGNUS runs and find that the E95 model tends to attribute more IR luminosity to AGN contribution, which results in less IR luminosity due to star formation -- hence, lower SFRs. This can also be seen in Figure \ref{agn_temp} as the E95 model shows a more significant component at longer wavelengths than the F06 and S12 models. This discrepancy between SFRs can influence the subsequent analysis. Based on the currently available data, it is not possible to decide which code produces the more accurate SFR estimates. We account for all five sets in the following analysis.

Total IR luminosity is the parameter estimated most consistently among the various SED codes and AGN models. This is expected as our \textit{Herschel} data points sample the peak of the IR spectrum, indicating that FIR observations are of vital importance in studying HLIRGs. We selected a subsample of HLIRGs with an AGN fraction $>0.1$ and reduced $\chi^2<5$ in SED fitting to define a clean and safe AGN subsample in Figure \ref{compare_lagn}. The AGN fraction requirement is due to the fact that AGN luminosity is only well constrained when AGN fraction has reached a large enough value. In terms of AGN luminosity, CYGNUS run using the E95 model predicts slightly higher AGN luminosity than all the other models. This also explains the larger dispersion of the long tail in the SFR comparison: similar total IR luminosity but higher AGN luminosity estimates naturally lead to lower estimates of IR luminosity due to star-formation activity.

The mean difference and 1-$\sigma$ scatter of estimates in the various SED-fitting runs with different AGN models are listed in Table \ref{difference}. In the AGN luminosity subtable, we only include  HLIRGs satisfying AGN fraction $>0.1$ and reduced $\chi^2<5$.
\begin{table*}
\caption{Median difference and 1-$\sigma$ scatter between various combinations of models. Each value is calculated by the estimates derived using the model in the given row divided by the estimates derived from the model in the given column. All are in a logarithmic scale (in unit of dex).}
\label{difference}      
\centering
\begin{tabular}{c c c c c c}
\hline\hline
\multicolumn{6}{c}{Stellar mass}\\
\hline
&CIGALE F06&CIGALE S12&CYGNUS E95&CYGNUS F06&CYGNUS S12\\
CIGALE F06&&-0.005$_{0.148}^{0.090}$&0.071$_{0.256}^{0.257}$&0.128$_{0.286}^{0.273}$&0.064$_{0.245}^{0.247}$\\
CIGALE S12&&&0.107$_{0.239}^{0.232}$&0.139$_{0.262}^{0.290}$&0.097$_{0.246}^{0.219}$\\
CYGNUS E95&&&&0.012$_{0.074}^{0.164}$&0.000$_{0.091}^{0.074}$\\
CYGNUS F06&&&&&-0.011$_{0.172}^{0.066}$\\
\hline
\multicolumn{6}{c}{SFR}\\
\hline
CIGALE F06&&-0.015$_{0.187}^{0.038}$&-0.018$_{0.199}^{0.529}$&-0.058$_{0.208}^{0.159}$&-0.095$_{0.193}^{0.128}$\\
CIGALE S12&&&0.013$_{0.137}^{0.564}$&-0.018$_{0.138}^{0.188}$&-0.042$_{0.133}^{0.099}$\\
CYGNUS E95&&&&-0.016$_{0.292}^{0.051}$&-0.020$_{0.506}^{0.042}$\\
CYGNUS F06&&&&&-0.005$_{0.128}^{0.042}$\\
\hline

\multicolumn{6}{c}{Total IR luminosity}\\
\hline
&CIGALE F06&CIGALE S12&CYGNUS E95&CYGNUS F06&CYGNUS S12\\
CIGALE F06&&-0.027$_{0.107}^{0.050}$&-0.096$_{0.166}^{0.110}$&-0.048$_{0.120}^{0.085}$&-0.039$_{0.115}^{0.076}$\\
CIGALE S12&&&-0.048$_{0.189}^{0.097}$&-0.012$_{0.115}^{0.096}$&0.004$_{0.108}^{0.072}$\\
CYGNUS E95&&&&0.018$_{0.033}^{0.158}$&0.021$_{0.029}^{0.165}$\\
CYGNUS F06&&&&&0.004$_{0.037}^{0.063}$\\
\hline

\multicolumn{6}{c}{AGN luminosity}\\
\hline
&CIGALE F06&CIGALE S12&CYGNUS E95&CYGNUS F06&CYGNUS S12\\

CIGALE F06&&-0.041$_{0.184}^{0.213}$&-0.335$_{0.394}^{0.358}$&-0.156$_{0.269}^{0.222}$&-0.042$_{0.255}^{0.249}$\\
CIGALE S12&&&-0.302$_{0.425}^{0.281}$&-0.117$_{0.277}^{0.203}$&-0.028$_{0.155}^{0.184}$\\
CYGNUS E95&&&&0.169$_{0.209}^{0.158}$&0.145$_{0.148}^{0.452}$\\
CYGNUS F06&&&&&0.016$_{0.120}^{0.351}$\\
\hline
\end{tabular}
\end{table*} 


In the following subsections, we analyze the nature of our HLIRG sample.  First we study their stellar masses, investigating their co-moving volume density at different redshifts and their contribution to the cosmic stellar mass density. Then we focus on their star-formation activity, mainly on their location relative to the star-forming main sequence (MS) and their contribution to the cosmic SFR density. Next we investigate their AGN activity by studying the BH growth rate as a function of redshift and stellar mass. Finally, we concentrate on the relationship between star-forming activity and BH growth in order to test whether the BH-galaxy co-evolution theory still holds for these extreme sources.

\subsection{Stellar mass properties}\label{mass}
\begin{figure}
\includegraphics[width=\linewidth]{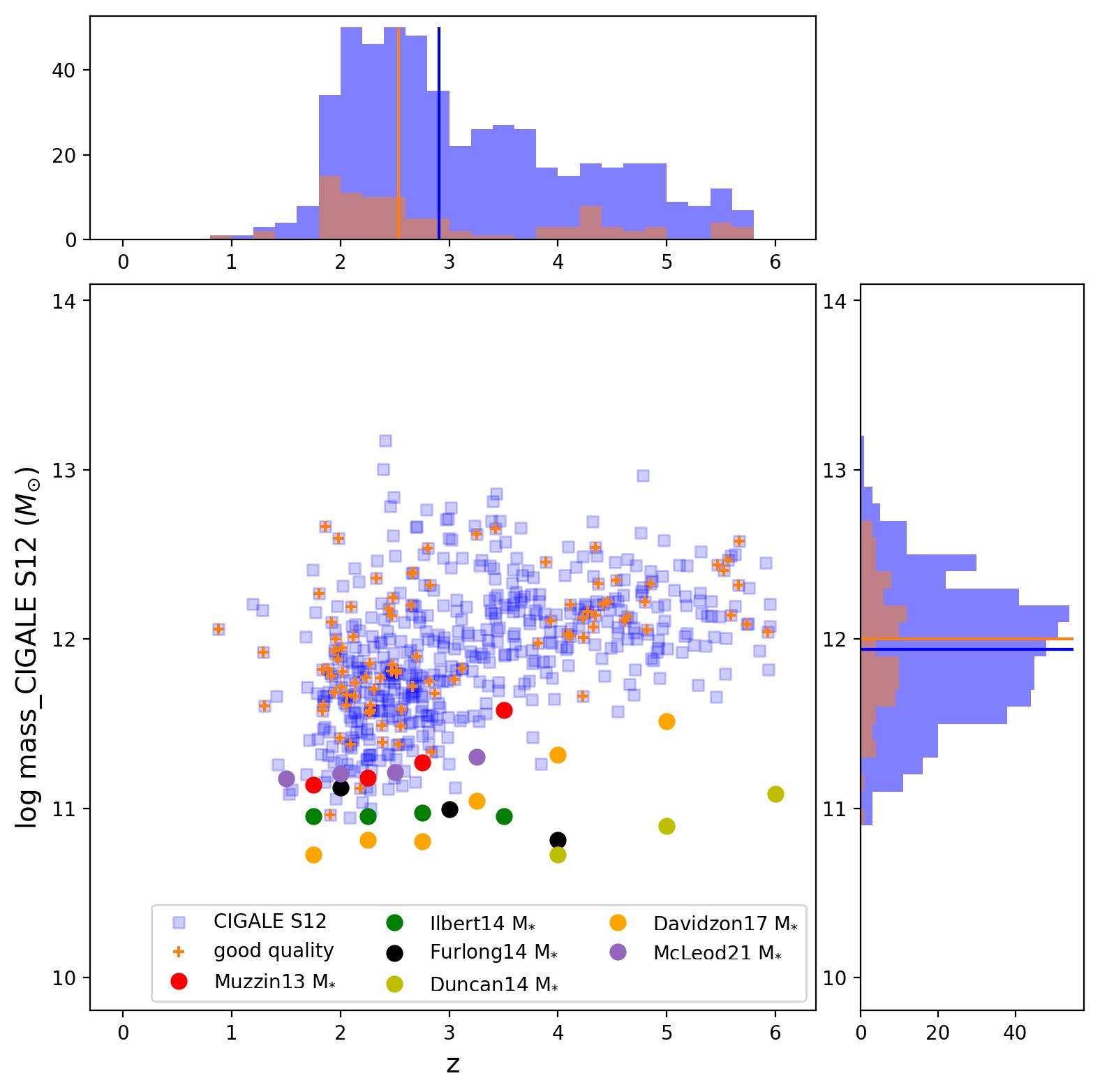}
\caption{Distribution of stellar mass estimates derived using CIGALE run with the S12 model as a function of redshift. The histograms show the distribution of both values and black solid lines indicate the median value of stellar mass estimates and redshift. We also add characteristic stellar mass values from various studies of the global stellar mass functions. }
\label{margin}
\end{figure}

In Figure \ref{margin}, we show stellar masses derived from CIGALE run using the S12 model versus redshift and the marginalized distributions of both parameters. Our HLIRGs reside in a wide redshift range, from $z\sim1$ to $z\sim6$. The bulk of the HLIRGs lies within $2<z<4$, peaking at $z\sim2,$ which coincides with the peak of the cosmic SFR density \citep[e.g.,][]{2014ARA&A..52..415M}, but with a significant tail towards higher redshift. Stellar mass estimates are between  $10^{11} M_{\odot}$ and  $10^{13} M_{\odot}$\citep[most massive galaxy nearby;][]{2010ApJ...715L.160C}, with a median value around $10^{12} M_{\odot}$, suggesting that our HLIRGs are a good sample for studying ultra-massive galaxies in the early Universe. We also plot the characteristic mass scales as a function of redshift from past studies of the global stellar mass function. It is clear that our HLIRGs are typically well above the characteristic mass scales at all redshifts. 

We ran a sanity check to validate the reliability of such ultra-massive HLIRGs. We first select HLRGs that have good fits, requiring reduced $\chi^2<5$ in results derived from all SED fitting codes and AGN models. We apply a quality cut on photometric redshifts for HLIRGs that have photometric redshifts, defined as (phot-z$_{max}$-phot-z$_{min}$)/2/(1+phot-z)<0.15, with phot-z$_{max}$ and phot-z$_{min}$ representing the lower and upper bound of the primary 80\% highest probability density (HPD) credible interval (CI) peak. In addition, we require reduced $\chi^2<5$ in the fit from the photometric redshift estimation code. We also exclude HLIRGs that have photometric redshift $z>4$ and significant $u$ band selection (>2-sigma) as they are unlikely to exist considering intergalactic medium (IGM) absorption. In the end, we selected a subsample of 93 HLIRGs, with 6 of them having spectroscopic redshifts and the remaining 87 having photometric redshifts. Their stellar mass versus redshift distribution is represented as orange crosses in Figure \ref{margin}. These HLIRGs, with good-quality photo-z estimates and SED fits, span a similar stellar mass range as the entire sample, with a similar median value around $10^{12} M_{\odot}$. 
The comparisons in Figures \ref{compare_mass}, \ref{compare_sfr}, \ref{compare_lir}, and \ref{compare_lagn}, along with the following analysis, do not change much either. We still focus on the entire HLIRG sample in the main text. We further discuss the robustness of our stellar mass estimates and show some examples of ultra-massive HLIRGs at high redshifts in Section \ref{mass_test_t} of the appendix.

Another possibility of such ultra-massive (and extremely luminous) HLIRGs is the lensing magnification effect, especially for galaxies at high redshifts. Magnified fluxes due to gravitational lensing will result in magnified luminosity and stellar mass. \citet{2010Sci...330..800N} found that bright submillimeter galaxies at $z>1$ that have 500 $\mu$m flux densities above 100 mJy are strongly lensed. In \citet{2021A&A...648A...8W}, we find that only $<1\%$ of our parent sample have 500 $\mu$m flux densities above 80 mJy. In our HLIRG sample, there are only three HLIRGs that have such high flux densities at 500 $\mu$m beyond $z>1$. Therefore, we conclude that our HLIRG sample is not significantly affected by contamination from lensed normal galaxies.

We investigate the co-moving volume density of our HLIRGs at different redshifts and compare with global stellar mass functions in observational studies \citep{2011ApJ...735L..34G, 2013ApJ...777...18M, 2013A&A...556A..55I, 2014MNRAS.444.2960D, 2017A&A...605A..70D, 2021MNRAS.503.4413M} and simulations \citep{2015MNRAS.450.4486F}. These observational studies are based on UV \citep{2011ApJ...735L..34G} or optical-NIR \citep{2013ApJ...777...18M, 2013A&A...556A..55I, 2014MNRAS.444.2960D, 2017A&A...605A..70D, 2021MNRAS.503.4413M} selections. The Evolution and Assembly of Galaxies and their Environment (EAGLE) simulations used in \citet{2015MNRAS.450.4486F} also tuned their sub-grid physics to match the observed stellar mass growth. All these literature mass functions are converted to the Salpeter IMF that we used in this study.

\begin{figure*}[htbp]
\includegraphics[width=\linewidth]{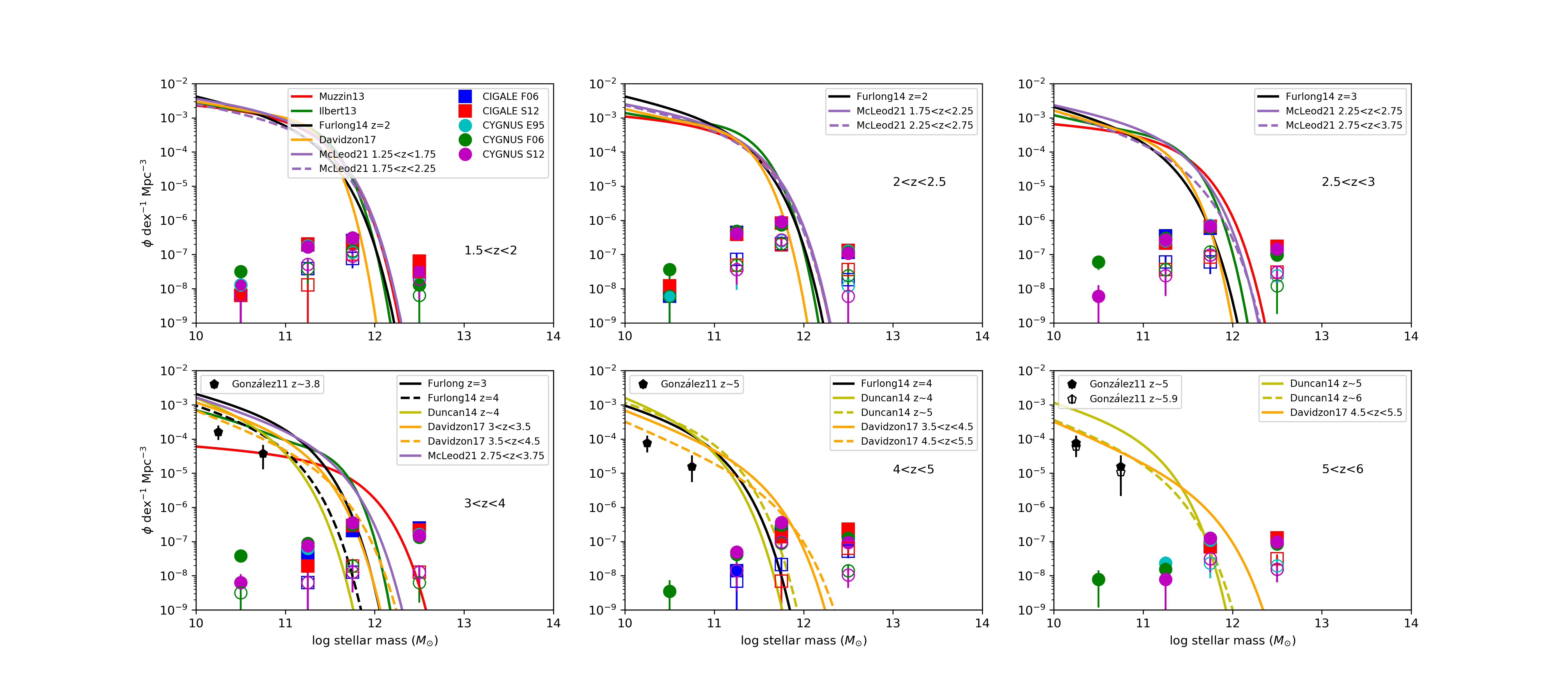}
\caption{Co-moving volume density of our HLIRGs in different redshift ranges compared with global stellar mass function measurements from the literature based on observations (red, green, yellow, orange, and purple lines as well as black pentagons) or simulations (black solid line). Error bars combine the Poisson error and the standard deviation among the three deep fields. Empty markers represent a subsample with good quality in the photo-z estimates and SED fits.}
\label{massfunc}
\end{figure*}

\begin{figure}
\includegraphics[width=\linewidth]{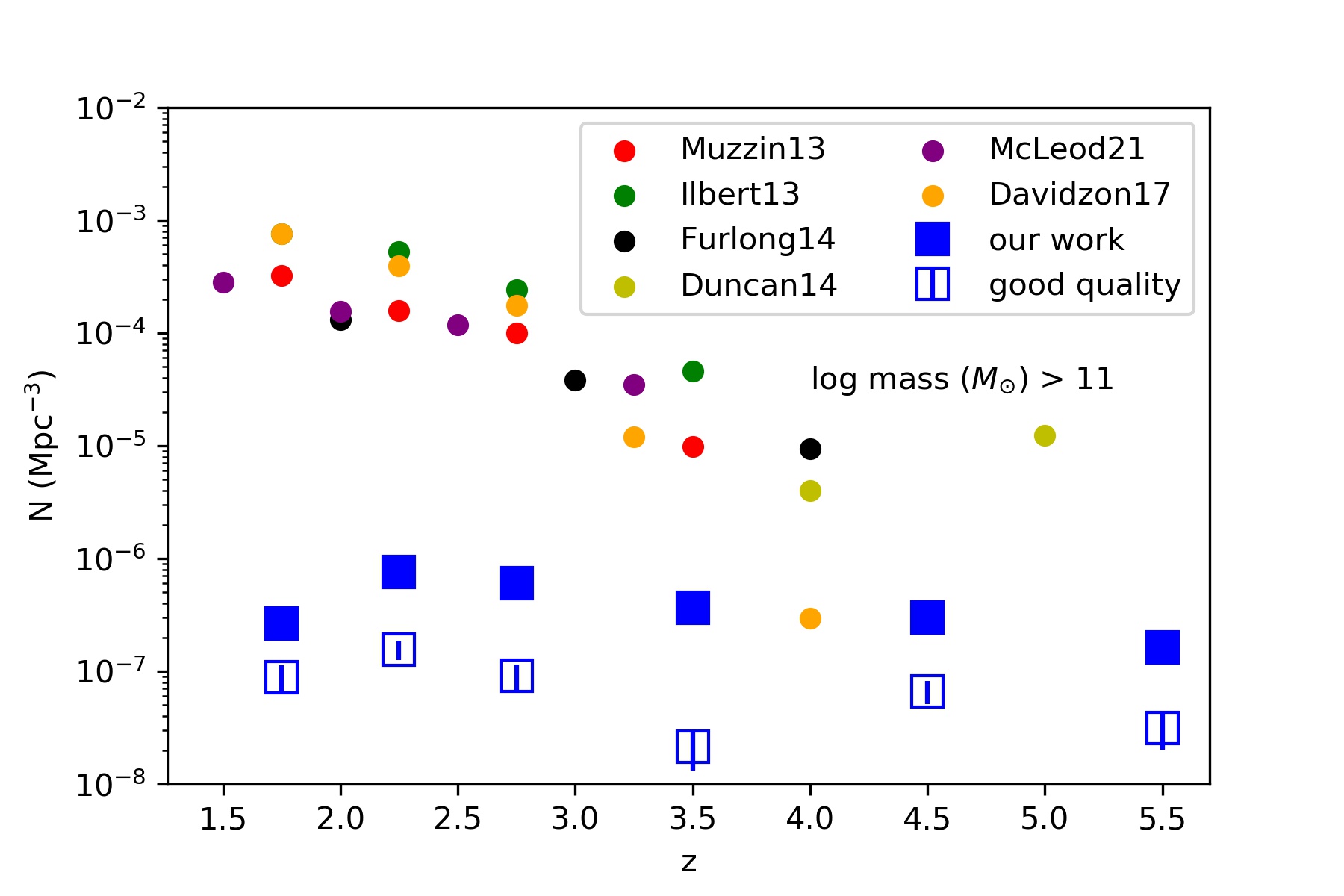}
\caption{Number density of ultra-massive galaxies above $10^{11} M_{\odot}$ as a function of redshift range compared with previous results in the literature. We use the mean value and standard deviation of number densities derived from the five models. Empty markers represent a subsample with good quality in the photo-z estimates and SED fits. }
\label{densityn11}   

 \end{figure}

Figure \ref{massfunc} shows that at the most massive end, the volume density of our HLIRGs is much higher than what has been reported in previous studies of global stellar mass functions. And these high volume densities of ultra-massive HLIRGs still hold when only considering the subsample with good-quality photo-z estimates and SED fits. This implies that the HLIRGs may be a dominant population in ultra-massive galaxies which have been largely missed in previous surveys. At $3<z<4$, observations and simulations show a large discrepancy at the most massive end. For example, \citet{2013ApJ...777...18M} reported a larger number density of massive galaxies. We integrate each global stellar mass functions above $10^{11} M_{\odot}$ to study the number density of massive galaxies at different epochs, and then compare them to the number density of our HLIRGs with the same mass threshold. As Figure \ref{densityn11} shows, our HLIRGs become more important in terms of the total number density of massive galaxies as redshift increases. A consistent picture can be found in \citet{2016ApJ...827L..25M}, in which above $z\sim2$, dusty star-forming galaxies dominate the galaxy population with masses $>10^{10.3} M_{\odot}$. Our sample demonstrates that dusty star-forming galaxies play an important role in the study of galaxy assembly in the most massive population in the early Universe \cite[see review][and references therein]{2014PhR...541...45C}.

We also study the contribution of our HLIRG sample to the cosmic stellar mass density. We adopt a widely used $1/V_{max}$ method \citep{1968ApJ...151..393S} to derive the total mass in each redshift bin. The co-moving volume $V_{max}$ for each source in one certain redshift bin is calculated by $V_{max}=V_{zmax}-V_{zmin}$, where $zmax$ is the minimum value between the upper boundary of that redshift bin and the maximum redshift at which the source can still be included in our study under the flux limit used to build our parent sample (i.e., \textit{Herschel} 250 $\mu$m flux of 35, 40, 45 mJy for EN1, LH, and Bo$\rm \ddot{o}$tes, respectively). We note here that we only include 388 HLIRGs that are from the unique sample and 66 HLIRGs that are from the multiple sample which means one \textit{Herschel} source has more than one counterparts in the LOFAR catalog. The remaining 72 multiple sources no longer satisfy our selection criteria because their fluxes fall below the flux cut after de-blending.

The left panel of Figure \ref{csfh} shows the evolution of the stellar mass density for our HLIRGs, compared with total galaxy mass density up to $z\sim 4$ in \citet{2013ApJ...777...18M} and $z\sim5$ in \citet{2021MNRAS.503.4413M, 2017A&A...605A..70D} as well as \citet{2010ApJ...716L.103L, 2011ApJ...735L..34G} at higher redshifts. All runs with different models produce similar stellar mass densities as a function of redshift, which first rises towards high redshift, then plateaus around $z\sim3$, and slightly declines out to $z\sim6$. Since cosmic mass density decrease as redshift increases, our HLIRGs contribute more to the cosmic stellar mass density at earlier epochs, increasing from $<0.1\%$ at $z\sim1$ to $\sim4-5\%$ at $z\sim5-6$.

One explanation of these ultra-massive HLIRGs we find in our study is the survey size. Our study is based on three deep fields that covers a wide area, reaching 26.65 deg$^2$ in total. The space density of ultra-massive HLIGRs above $10^{12} M_{\odot}$ is 6-9 (obtained from CIGALE and CYGNUS runs respectively) galaxies/deg$^2$. The number reduces to 1-2 galaxies/deg$^2$ if we consider the good-quality subsample. Previous studies of global stellar mass function covered much smaller areas: \citet{2013ApJ...777...18M}, \citet{2013A&A...556A..55I} and \citet{2017A&A...605A..70D} selected their sample from COSMOS filed (2 deg$^2$); \citet{2014MNRAS.444.2960D} studied galaxies in CANDEL/GOODS-S field (170 arcmin$^2$); \citet{2021MNRAS.503.4413M} was based on a raw survey area of 3 deg$^2$. At most a few ultra-massive galaxies above $10^{12} M_{\odot}$ can be included in such a small area. This demonstrates that a large survey size is important in building a large sample of ultra-massive galaxies to benefit the study of galaxy evolution especially at early epoch. 

In summary, our HLIRG sample indicates that potentially there are  many more ultra-massive galaxies that have been missed in surveys based on UV or optical-NIR data. In addition, HLIRGs contribute more to the cosmic stellar mass density towards higher redshifts, from $<0.1\%$ at $z\sim1$ to $\sim4-5\%$ at $z\sim5-6$.

\subsection{Star-forming activity}

\subsubsection{Location on the star-forming main sequence}
Many studies have found a tight correlation ($\sim 0.3$ dex) between galaxy stellar mass and SFR, which has been dubbed the star-forming "main sequence" (MS) \citep{2004MNRAS.351.1151B, 2007ApJ...660L..43N, 2007A&A...468...33E}. This relation extends to high redshifts with general trends showing that at a given stellar mass, SFR increases with redshift \citep{2007ApJ...670..156D, 2011ApJ...739L..40R, 2012ApJ...754L..29W, 2010ApJ...714.1740M, 2014ApJ...781...34G, 2013MNRAS.431..648W, 2015ApJ...799..183S, 2014ApJ...791L..25S,2018A&A...615A.146P}, probably owing to higher gas fraction \citep{2013ApJ...768...74T}. We adopt the \citet{Speagle} correlation which combined 64 MS relations in 25 studies with redshift ranging from 0 to 6, providing a time-dependent form of MS, to investigate the locations of HLIRGs with relative to MS (represented by $\Delta MS$=SFR-SFR$_{MS}$). 

\begin{figure*}[htbp]
\includegraphics[width=\linewidth]{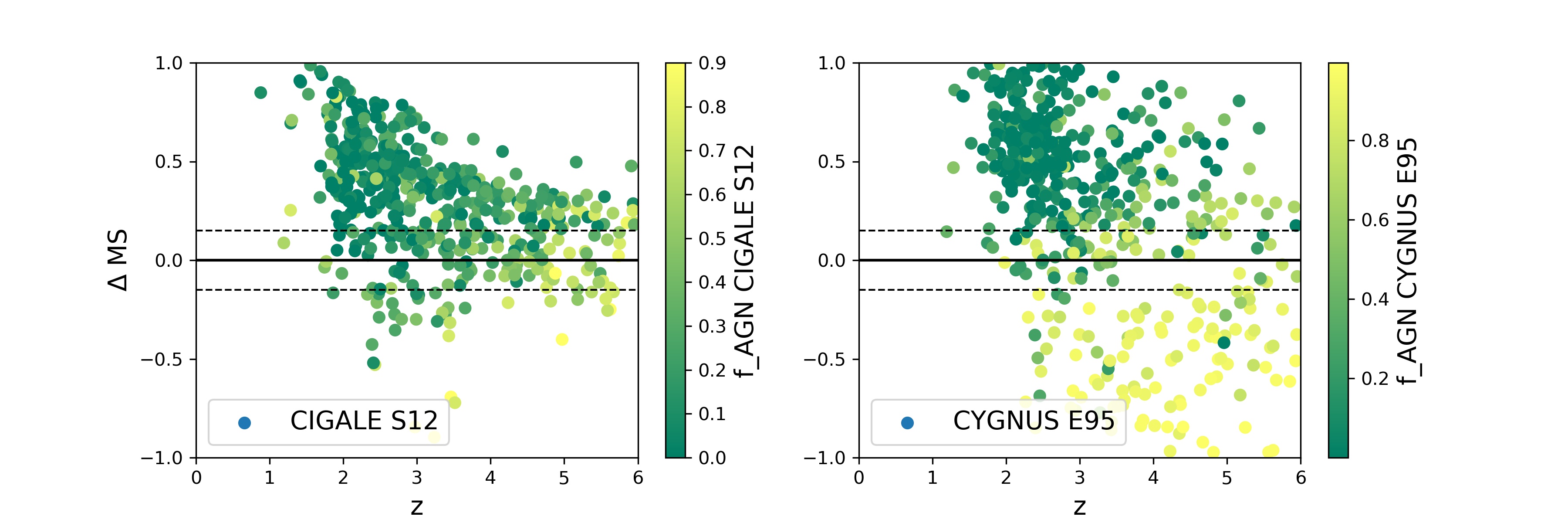}
\caption{Distance to MS as a function of redshift for results from CIGALE run with the S12 model and CYGNUS run with the E95 model, color-coded by AGN fraction estimates. The solid and dashed lines are the location of the MS as determined in \citet{Speagle} and the widely adopted 0.3 dex width of the MS, respectively.}
\label{ms_loc}
\end{figure*}

\begin{figure*}[htbp]
\resizebox{\hsize}{!}{\includegraphics[width=\linewidth]{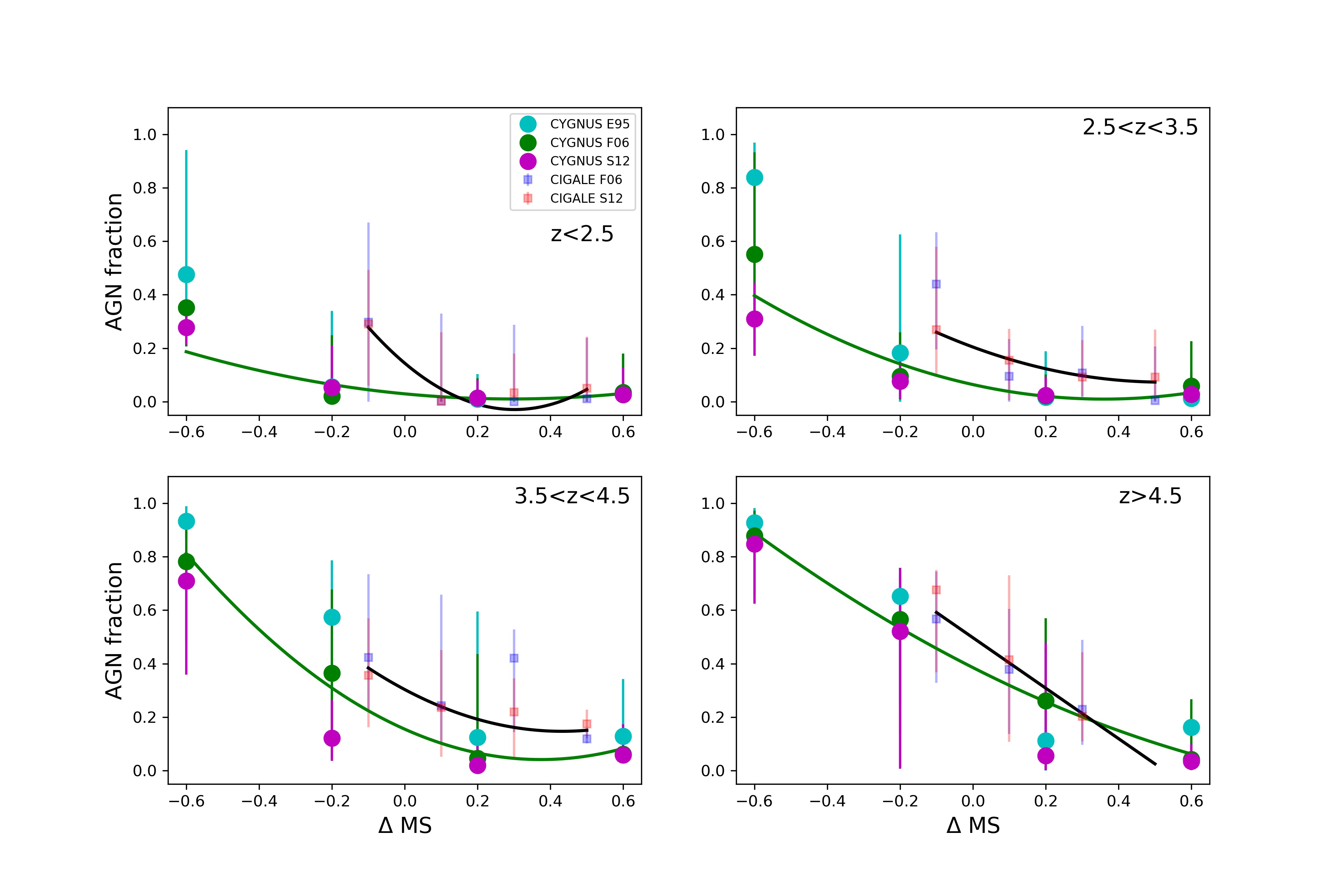}}
\caption{Median AGN fraction estimates of CIGALE and CYGNUS results as a function of $\Delta MS$ in different redshift bins. The error bars indicate the upper and lower quartile within each $\Delta MS$ range. The black and green lines are polynomial fitting combing the two CIGALE runs and the three CYGNUS runs, respectively.}
\label{f_agn_delta}
\end{figure*}

Figure \ref{ms_loc} shows the $\Delta MS$ as a function of redshift for CIGALE run with the S12 model (similar to the F06 model) and CYGNUS run with the E95 model (similar to the F06 and S12 models). We chose the CIGALE S12 run over the F06 run because it typically shows a better fitting performance when we visually inspect the best-fit templates. We chose the CYGNUS E95 run over the other two runs because it has the largest deviation when comparing CIGALE and CYGNUS (see AGN luminosity panel in Table \ref{difference}), which is helpful in demonstrating the differences when using different fitting methods. The majority of our HLIRG sample has a positive value of $\Delta MS$, meaning that most HLIRGs are actively star-forming or in the starburst regime. Those with large AGN contributions are more likely to have smaller or negative $\Delta MS$, suggesting they are undergoing suppressed star formation, which is consistent with an AGN quenching scenario. This trend is more obvious in results from CYGNUS run with the E95 model as 82\% of the HLIRGs lying below the MS have AGN fractions larger than 0.7. This may also be affected by the fact that the E95 model assigns more IR luminosity to AGN and less to IR luminosity due to star formation, which results in lower SFRs. However, the other two AGN models in the CYGNUS runs still show a similar trend, which is consistent with the AGN quenching scenario. The subsample of reliable phot-z estimates and SED fits shows a similar picture.

Figure \ref{f_agn_delta} shows the distribution of AGN fraction estimates in each $\Delta MS$ bin at different redshift ranges for all CIGALE and CYGNUS runs. We also plot the fits for the two CIGALE runs together and the three CYGNUS runs together, respectively. In all redshift ranges, both CIGALE and CYGNUS estimates are consistent with an AGN driven quenching phenomenon, with HLIRGs below the MS typically associated with large AGN fractions and HLIRGs above the MS typically associated with small AGN fractions. In addition, this trend becomes stronger towards higher redshifts. According to the best-fit lines, there may exit a weak rising tail toward the largest $\Delta MS$ bin except the highest redshift range, in which the CIGALE runs lack data and CYGNUS runs have only $< 5$ sources. However, we argue that the weak rising trend in the largest $\Delta MS$ bins is not statistically significant. We also look into the median stellar mass in each $\Delta MS$ bin, finding a decreasing trend as $\Delta MS$ bin increases (maximum >1 dex between the first and the last $\Delta MS$ bin). We cannot rule out the possibility that HLIRGs below the MS are also affected by other factors connected to large stellar mass, such as exhaust of gas supply or morphological quenching \citep[e.g., ][]{2009ApJ...707..250M}. In any case, our results do show a negative relationship between AGN activity and SFR and this negative relationship is more prominent towards high redshifts. These negative trends still hold in the subsample of reliable phot-z estimates and SED fits, but with a weaker redshift evolution due to poor statistics.

The connection between AGN activity and star-forming activity is still widely debated. There is observational and theoretical evidence that suggest that AGN can suppress the SFR \citep{2005Natur.433..604D, 2006MNRAS.365...11C, 2007MNRAS.380..877S, 2012A&A...537L...8C, 2012ApJ...745..178F, 2012Natur.485..213P} as well as evidence to the contrary \citep{2012ApJ...760L..15H, 2012MNRAS.427.3103B, 2013ApJ...772..112S}. In our study, we find a downward trend between the AGN fraction and distance to the MS, which may be consistent with an AGN quenching scenario. However, the main weakness of our study is that both AGN fractions and SFR estimates are derived from SED fitting to the same multi-wavelength photometry rather than derived from completely independent observations.

\subsubsection{Contribution to cosmic SFR density}
\begin{figure*}[htbp]
\resizebox{\hsize}{!}{\includegraphics[width=\linewidth]{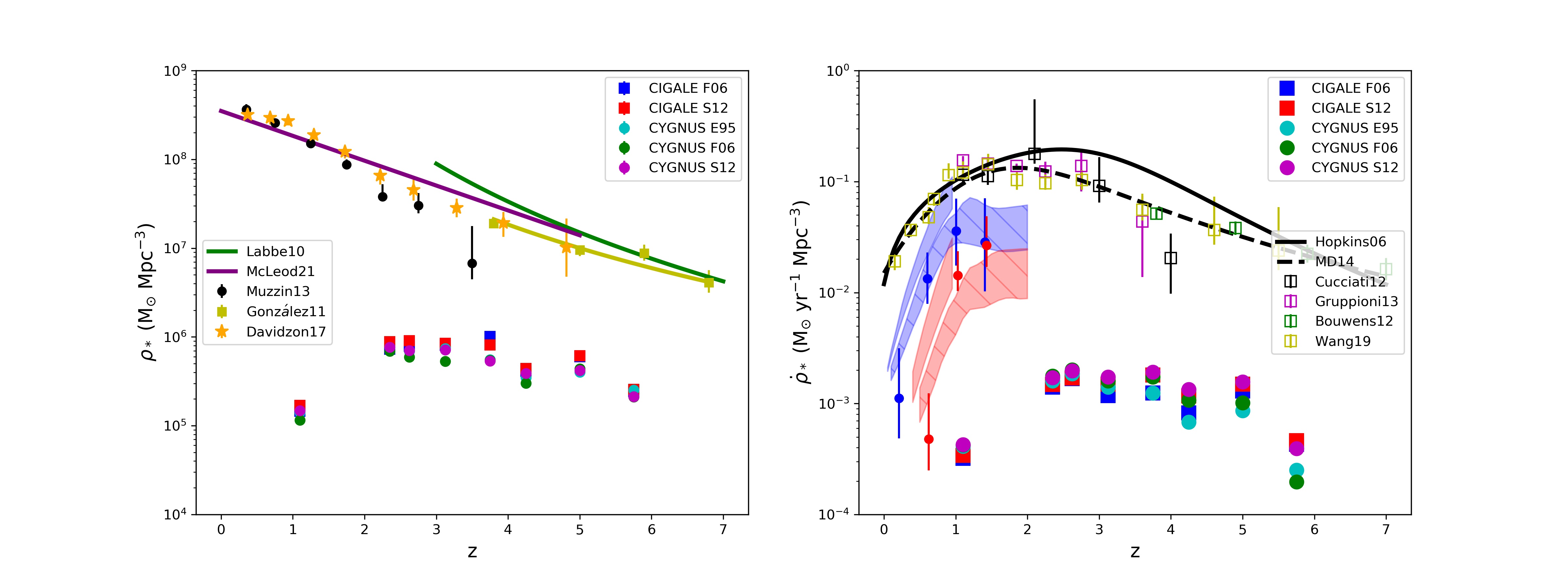}}
\caption{Contribution of our HLIRGs to the cosmic stellar mass density (left panel) and the cosmic SFR density (right panel). In the right panel, black solid and dashed lines are taken from \citet{2006ApJ...651..142H} and \citet{2014ARA&A..52..415M}, respectively. Squares represent various studies of the global SFR density. Blue and red area and circles are contribution from LIRGs and ULIRGs respectively, taken from \citet{2005ApJ...632..169L},\citet{2011A&A...528A..35M} (hatched area) and \cite{2012ApJ...761..140C} (circles). All data points are converted to the Salpeter IMF.}
\label{csfh}
\end{figure*}

In this subsection, we investigate the contribution of HLIRGs to the cosmic SFR density. We adopted the same $1/V_{max}$ method as Section \ref{mass} and only include 388 and 66 HLIRGs in the unique sample and the multiple sample, respectively. The right panel of Figure \ref{csfh} shows the contribution of our HLIRGs to the cosmic SFR density in different redshift ranges. We also add the contribution of LIRGs and ULIRGs from \citet{2005ApJ...632..169L, 2011A&A...528A..35M, 2012ApJ...761..140C}. Both \citet{2005ApJ...632..169L} and \citet{ 2011A&A...528A..35M} used 24 and 70 $\mu m$ selected sources and \citet{2012ApJ...761..140C} used \textit{Herschel} selected sources. It is clear that the total SFR density of HLIRGs peaks at $z\sim 2-3$, when the cosmic SFR also reaches its peak, then becomes flat and slightly drops at $z>5$. Similar to the cosmic mass density, since the cosmic SFR density shows a downward trend after its peak, the contribution from HLIRGs rises from $\sim0.5\%$ to $\sim 4\%$ at $z\sim5$ \citep[using black dashed line in][]{2014ARA&A..52..415M}. 

In summary, our HLIRG sample agrees with the AGN quenching scenario as AGN fraction increases with decreasing distance to the star-forming MS. This trend is stronger at higher redshift. Our sample confirms that toward higher redshift, HLRGs contributes more to the cosmic SFR density.

\subsection{AGN activity}
In this subsection, we investigate AGN activity in our HLIRG sample. We use the AGN bolometric luminosity to derive the BH growth rate using the following formula \citep[e.g.,][]{2012ApJ...753L..30M}:
\begin{equation}
\dot{M}_{BH}=\frac{(1-\epsilon)L_{bol}}{\epsilon c^2}
,\end{equation}
where $\epsilon$ is the efficiency by which the accreting mass is converted into energy. We adopted the widely used value of 0.1\citep{1982MNRAS.200..115S}, meaning that radiation emitted by AGN comes from the energy transferred from 10\% of its accreting material. Similar to AGN luminosity comparisons in Section \ref{comp}, we selected HLIRGs with their AGN fraction $>0.1$ and we required reduced $\chi^2\le5$ to build a clean and safe AGN sample.

\begin{figure}[htbp]
\resizebox{\hsize}{!}{\includegraphics[width=\linewidth]{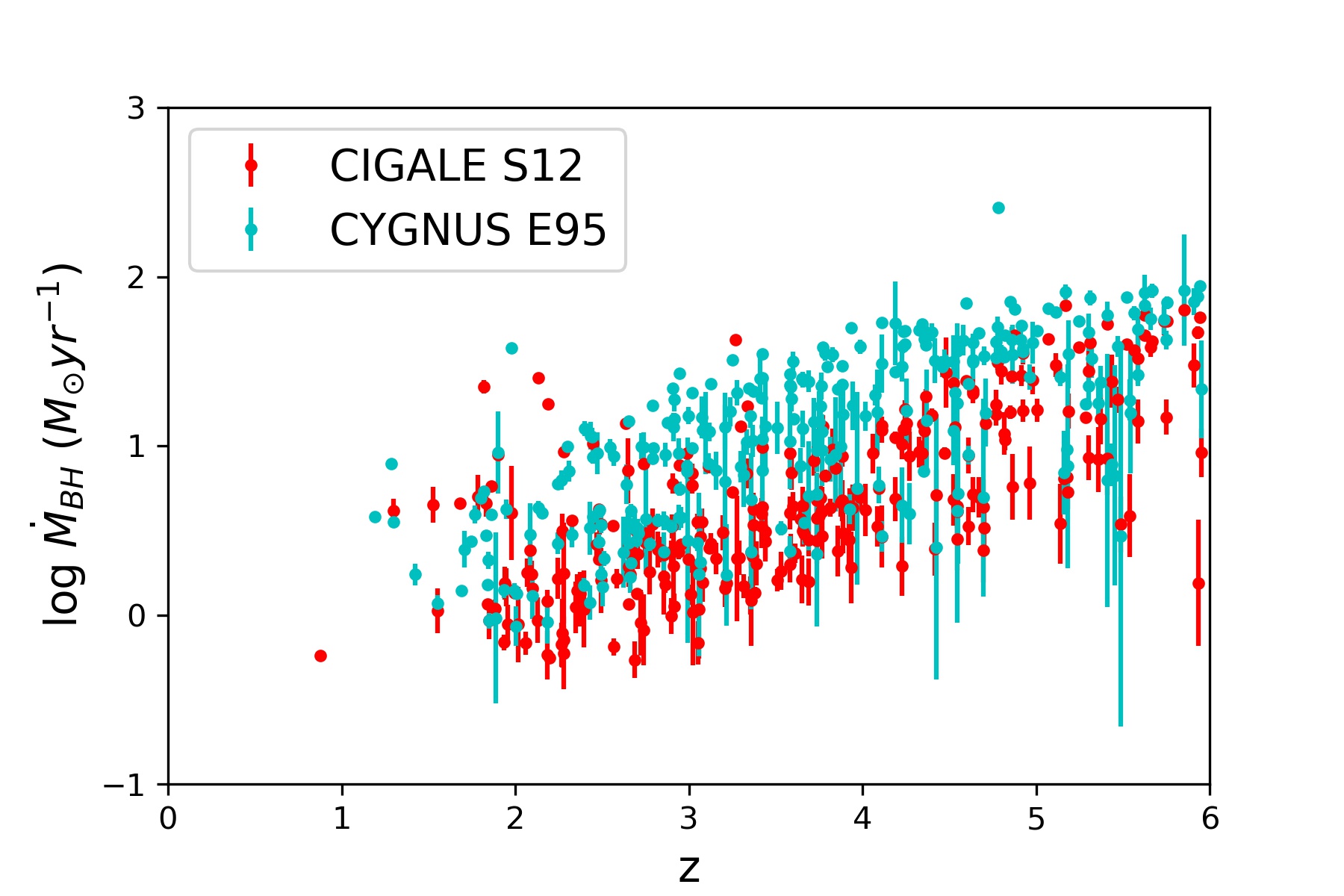}}
\caption{Distribution of BH growth rate versus redshift for our HLIRG sample derived from CIGALE run with the S12 model and CYGNUS run with the E95 model.}
\label{bha_z}
\end{figure}
Figure \ref{bha_z} shows the distribution of BH growth rate as a function of redshift from CIGALE run with the S12 model and CYGNUS run with the E95 model. The other three SED fitting runs (i.e., CIGALE run with F06 model, CYGNUS runs with F06 and S12 models) produce similar results. All results show an increasing BH growth rate as redshift increases. This is expected because our sample is flux-limited. Galaxies at earlier epochs would be easier to detect if they are experiencing more rapid BH growth and emit a larger amount of energy that is then absorbed and re-emitted by dust.

\begin{figure*}[htbp]
\resizebox{\hsize}{!}{\includegraphics[width=\linewidth]{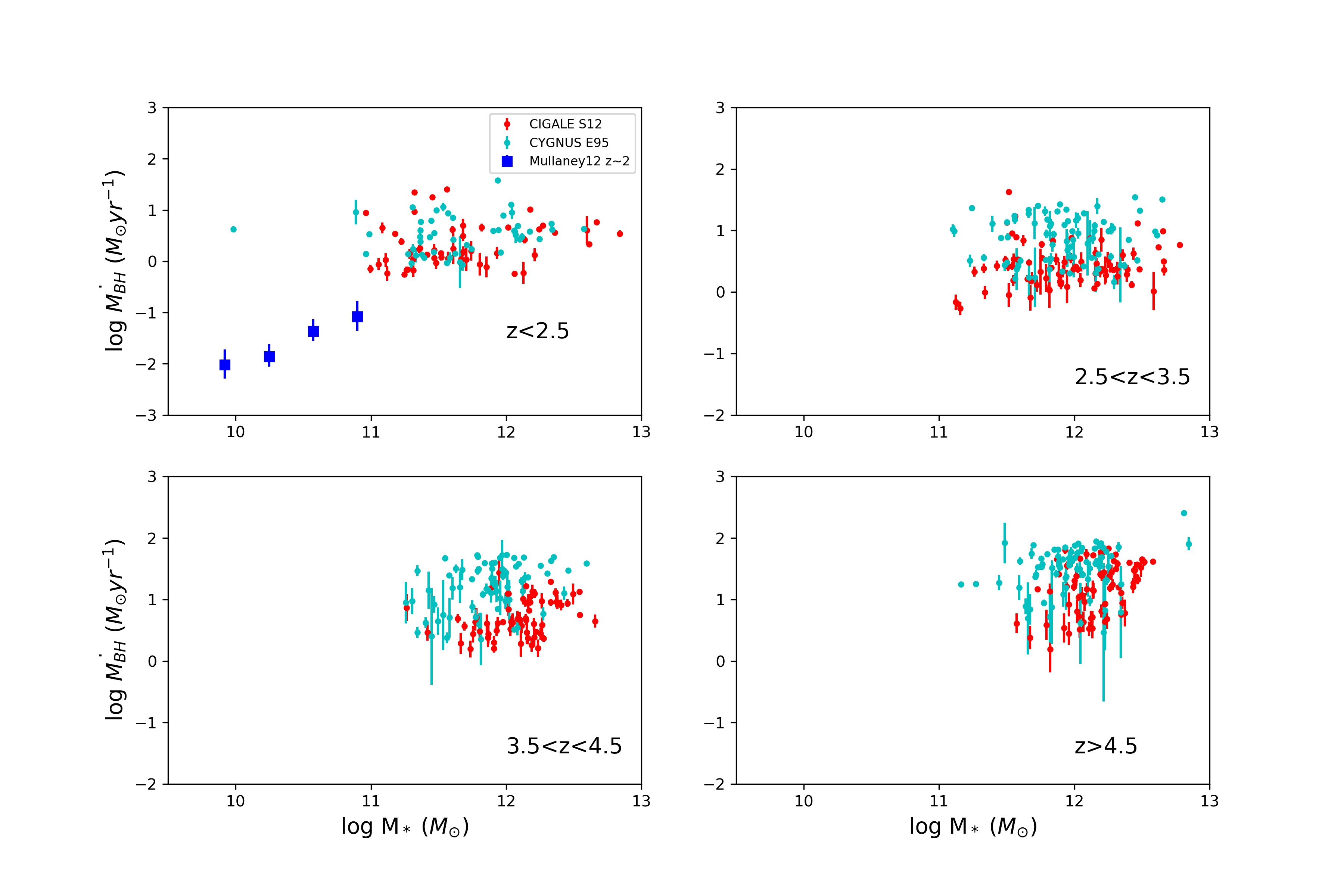}}
\caption{BH growth rate as a function of stellar mass over different redshift ranges. Blue squares are taken from \citet{2012ApJ...753L..30M} at $z\sim2$.}
\label{BHgrow_mass}
\end{figure*}
Figure \ref{BHgrow_mass} shows the BH growth as a function of galaxy stellar mass in different redshift ranges. The exact values of BH growth rate vary among different SED runs in each redshift range, however, neither of the five codes or models show any clear evolution of BH growth as a function of stellar mass. In the lowest redshift bin, we also make a comparison with the data from \citet{2012ApJ...753L..30M}. \citet{2012ApJ...753L..30M} studied a sample of SF galaxies at $z\sim1$ and $z\sim2$ and used X-ray stacking to derive average BH growth rate. They found an increasing trend of BH growth as stellar mass increases in both redshift bins. Our data probe a more massive range, which may imply that towards the massive end, the BH growth rate may reach a limiting value as stellar mass increases.

\subsection{SF-AGN co-evolution}
In this subsection, we study the correlation (if any) between the host galaxy SFR and its BH growth rate. We first study the SFR versus AGN luminosity distribution in different redshift ranges. Then we investigate the ratio of BH growth rate to SFR as a function of stellar mass in order to explore whether there exists any trends. Increasing or decreasing trends would suggest that as stellar mass increases, the central black holes grow relatively faster or slower than their host galaxy, which are imcompatible with the correlation between BH mass and host galaxy stellar mass for galaxies with different stellar mass. Flat trend supports that BH grows at a relative stable speed compared to the host galaxy across all stellar mass range, which brings up the correlation between BH mass and host galaxy mass.


\begin{figure*}[htbp]
\resizebox{\hsize}{!}{\includegraphics[width=\linewidth]{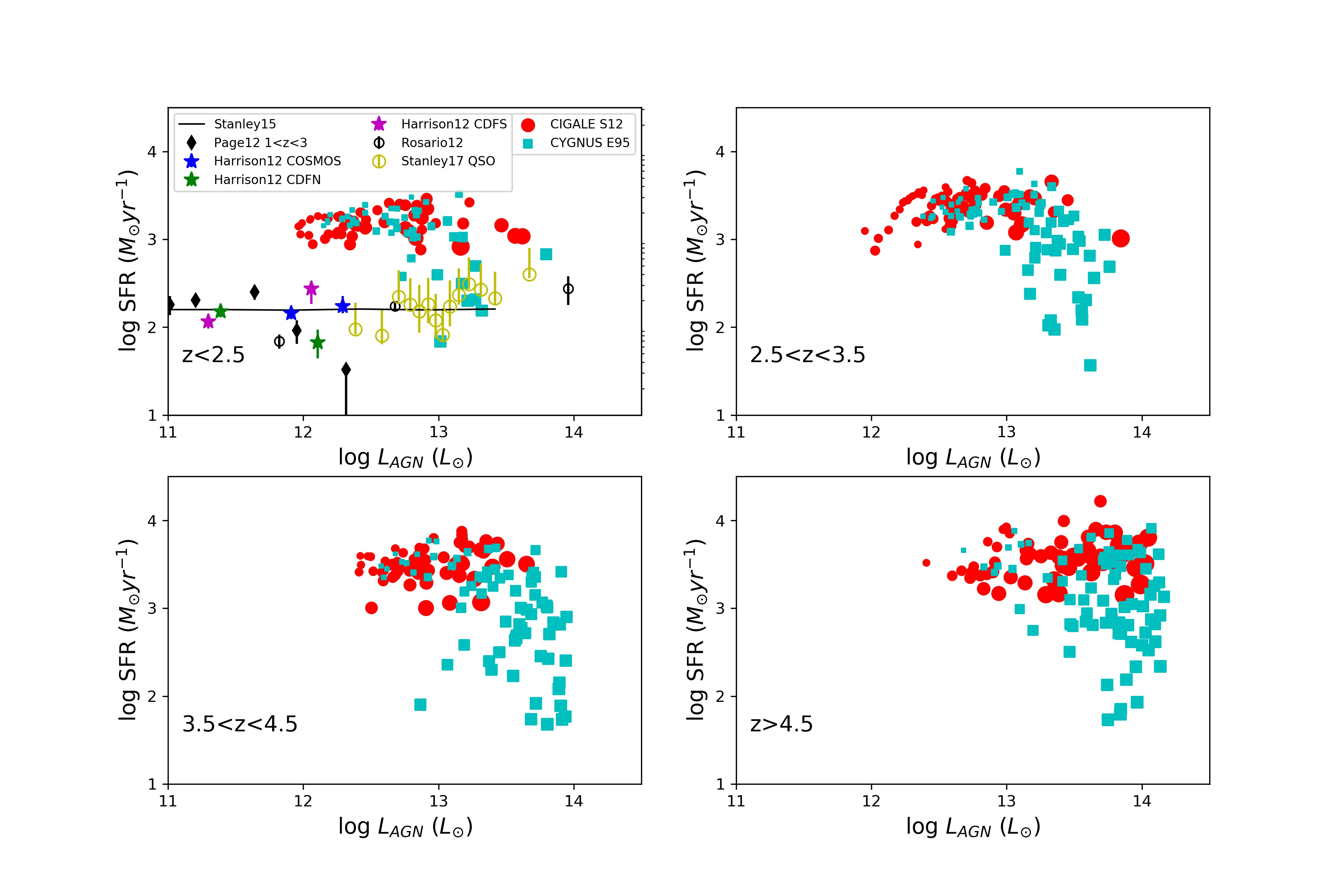}}
\caption{Correlation between SFR and AGN luminosity for estimates derived using CIGALE run with the S12 model and CYGNUS run with the E95 model in four different redshift bins that are size-coded by AGN fraction. We also show the previous work from \citet{2012ApJ...760L..15H, 2012Natur.485..213P, 2012A&A...545A..45R, 2015MNRAS.453..591S,2017MNRAS.472.2221S}.}
\label{sfr_lagn_z}
\end{figure*}

Figure \ref{sfr_lagn_z} shows the correlation between SFR and AGN luminosity derived using  CIGALE S12 and CYGNUS E95 SED fitting results in different redshift bins. CIGALE S12 estimates are more flat within each redshift bin, while CYGNUS E95 estimates show a large population of HLIRGs that have low SFRs when AGN luminosities exceed $10^{13} L_{\odot}$, consistent with an AGN quenching scenario in galaxies hosting most luminous AGNs. Similarly to Figure \ref{ms_loc}, this is also due to the fact that this model tends to attribute more IR luminosity to AGN contribution resulting in lower SFR estimates (see Section \ref{comp}). We also included data from \citet{2012ApJ...760L..15H, 2012Natur.485..213P, 2012A&A...545A..45R, 2015MNRAS.453..591S} that demonstrate a field-dependent trend (i.e., different trends in different fields), as well as a downward trend, upward trend, and flat trend, respectively. These studies utilized X-ray detected AGNs that have lower bolometric AGN luminosities than our HLIRGs. We also add data from more luminous QSO sample in \citet{2017MNRAS.472.2221S} which found a positive trend. Similarly to Figure \ref{f_agn_delta}, some degree of (anti-) correlation may exist between AGN luminosity and SFR estimates as they are both derived from fitting to the same multi-wavelength photometry. However, CIGALE and CYGNUS results show different trends. It is not possible with currently available data to determine whether it is CIGALE or CYGNUS that is correct or whether it is neither. We need more data in the longer wavelength range to differentiate and derive more accurate measurements of the AGN and SF contribution to the far-IR band.



\begin{figure*}[htbp]
\resizebox{\hsize}{!}{\includegraphics[width=\linewidth]{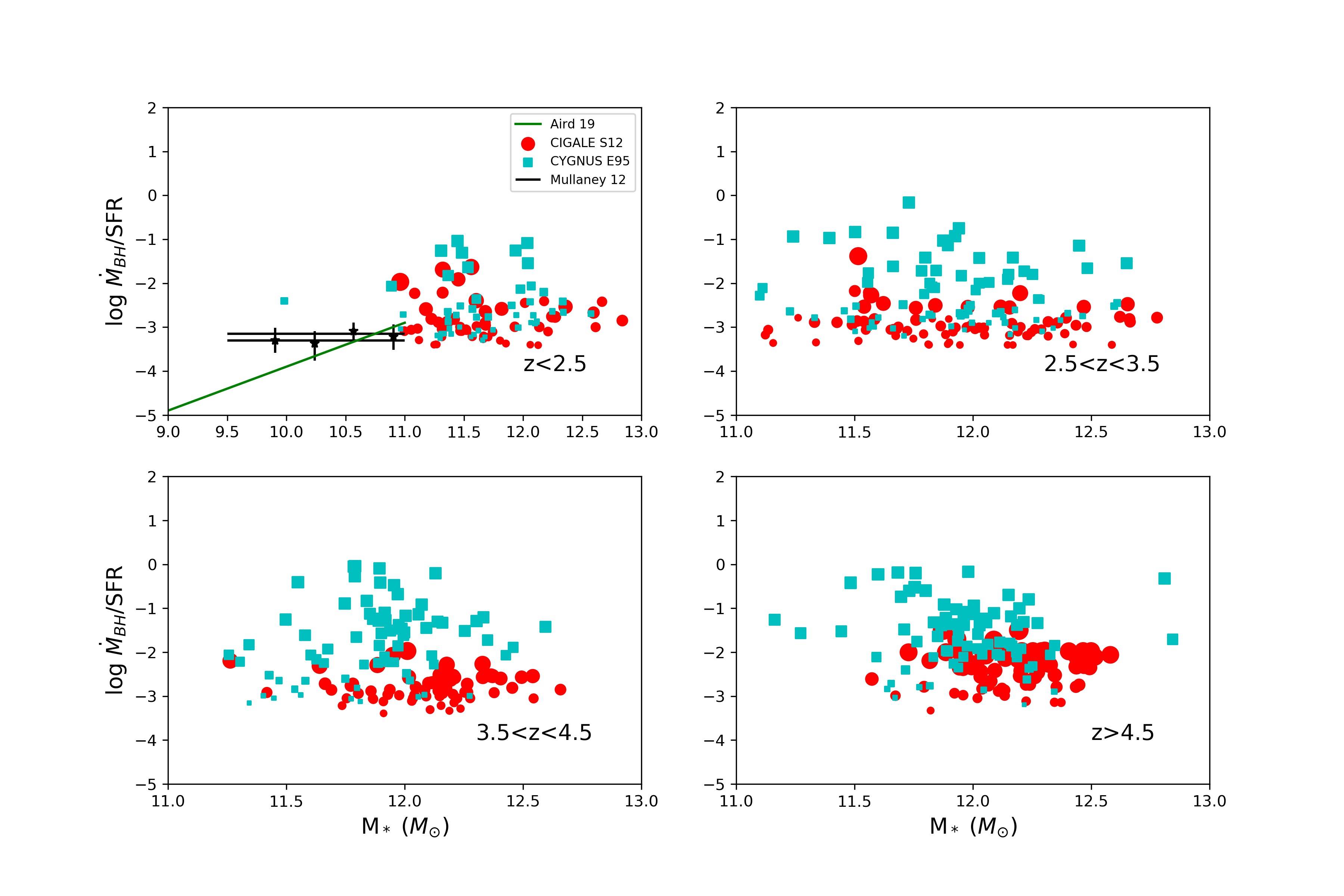}}
\caption{BH growth rate to SFR ratio as a function of stellar mass at different redshifts. The sizes indicate the AGN fractions. Black asterisks are from \citet{2012ApJ...753L..30M} and black solid lines are their fits. Green line is taken from \citet{2019MNRAS.484.4360A}}
\label{bh_sf_ratio}
\end{figure*}
Figure \ref{bh_sf_ratio} shows the BH growth to SFR ratio as a function of stellar mass at different redshifts. The CIGALE S12 results show that the BH growth to SFR ratio is nearly flat across the stellar mass range, similar to what has been found in \citet{2012ApJ...753L..30M} and consistent with the scenario of co-evolution between AGN and host galaxies. In contrast, \citet{2019MNRAS.484.4360A} revealed a positive trend in low-to-moderate mass galaxies. The results derived from CYGNUS E95 results are more scattered, especially for HLIRGs that have large AGN fractions at higher redshifts. This is also due to the fact that CYGNUS produces much smaller SFRs for these sources.


\section{Conclusions}\label{conclu}
In this paper, we present our sample of 526 HLIRGs with redshifts $1<z<6$ in the Bo$\rm \ddot{o}$tes, Lockman-Hole, and ELAIS-N1 fields. We adopted two SED fitting methods, CIGALE  -- based on energy balance -- and CYGNUS -- based on radiative transfer models and no assumption of energy balance. We used two and three different AGN models in CIGALE and CYGNUS, respectively. We compare their estimates and study the properties of our HLIRG sample, finding that:

\begin{itemize}
  \item There is no significant systematic offset between CIGALE and CYGNUS estimates for the majority of our HLIRGs in any of the galaxy properties we investigated (stellar mass, SFRs, IR luminosity, AGN luminosity). SFRs exhibit large dispersion that leads to different trends in the analyses presented in this paper.
  
  \item Our HLIRGs are ultra-massive, with a median mass of $10^{12} M_{\odot}$. There are potentially many more ultra-massive galaxies than found in previous studies based on UV or optical-NIR data and simulations. Our work suggests that as redshift increases, these HLIRGs contribute more to the cosmic stellar mass density.
  
 \item There is an anti-correlation between AGN fraction and the distance to star-forming MS ($\Delta$MS) in which high AGN fractions are more likely to be linked to galaxies that lie below the MS, while low AGN fractions are usually associated with galaxies that lie above the MS, which is consistent with the AGN quenching scenario. Similar to stellar mass density, HLIRGs contributes more to the cosmic SFR density as redshift increases.
 
 \item There is a flattening trend of BH growth rate as stellar mass increases, implying that our HLIRG sample may reach a maximum value of BH growth rate. Due to the large dispersion of SFR estimates between CIGALE and CYGNUS runs, we observe differences in the relationship between SFR and AGN luminosity. CIGALE shows a relatively flat trend at all redshifts while CYGNUS finds that SFR decreases as AGN luminosity increases, which agrees with AGN quenching. This large dispersion of SFR estimates also causes differences in BH growth rate to SFR ratio as a function of stellar mass. CIGALE runs exhibit a relatively flat trend, consistent with the scenario in which BH co-evolves with its host galaxy, while runs with CYGNUS exhibit a more scattered distribution. These differences, resulting from the methods used, may partly explain the contrasting results found in some of the previous studies. 
 
Our study finds that HLIRGs are ultra-massive, extremely dusty star-forming, and experiencing very active black hole accretion. They are essential to improving the understanding of galaxy evolution in extreme conditions, especially at high redshifts. In the future, we need more spectroscopic observations to confirm their redshifts, their hyper luminous and ultra-massive nature, as well as to study their gas content and gas kinematics, etc. We also need more photometric data in the longer wavelength range to better distinguish the contribution from star-formation activity and black hole accretion.
  
\end{itemize}
\begin{acknowledgements} 
We acknowledge support from UK Science and Technology Facilities Council (STFC) via grant ST/R505146/1, grant ST/R000972/1 and grant ST/R504737/1, INAF under the SKA/CTA PRIN “FORECaST” and the PRIN MAIN STREAM “SAuROS” projects, from the Ministero degli Affari Esteri e della Cooperazione Internazionale - Direzione Generale per la Promozione del Sistema Paese Progetto di Grande Rilevanza ZA18GR02, and from the Polish National Science Centre via grant UMO-2018/30/E/ST9/00082.
\end{acknowledgements}
\bibliographystyle{aa}
\bibliography{hlirgbib}

\begin{appendix}

\section{Examples of best-fit SEDs}
\begin{figure*}
\centering
        \begin{subfigure}{0.3\textwidth}
                \includegraphics[width=\textwidth]{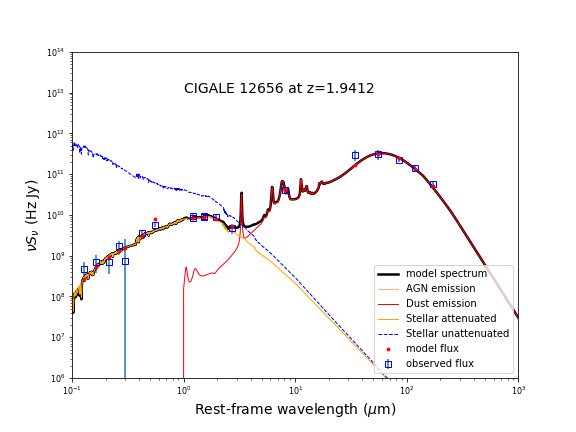}
        \end{subfigure}
        \begin{subfigure}{0.3\textwidth}
                \includegraphics[width=\textwidth]{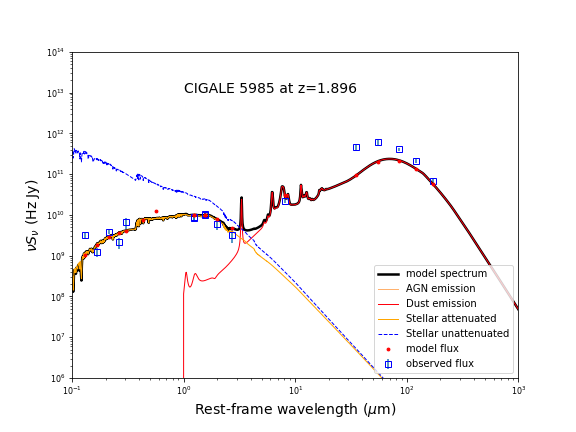}
        \end{subfigure}
        \begin{subfigure}{0.3\textwidth}
                \includegraphics[width=\textwidth]{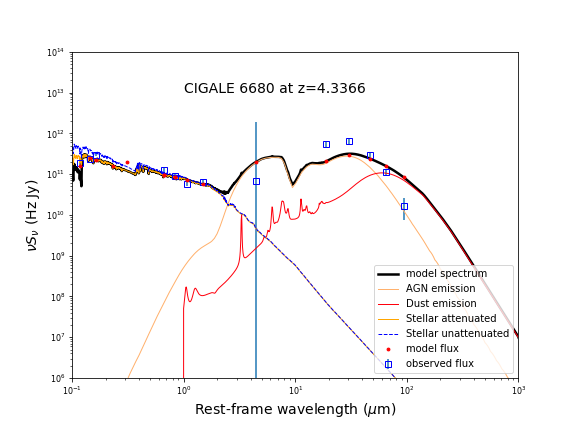}
        \end{subfigure}
        \begin{subfigure}{0.3\textwidth}
                \includegraphics[width=\textwidth]{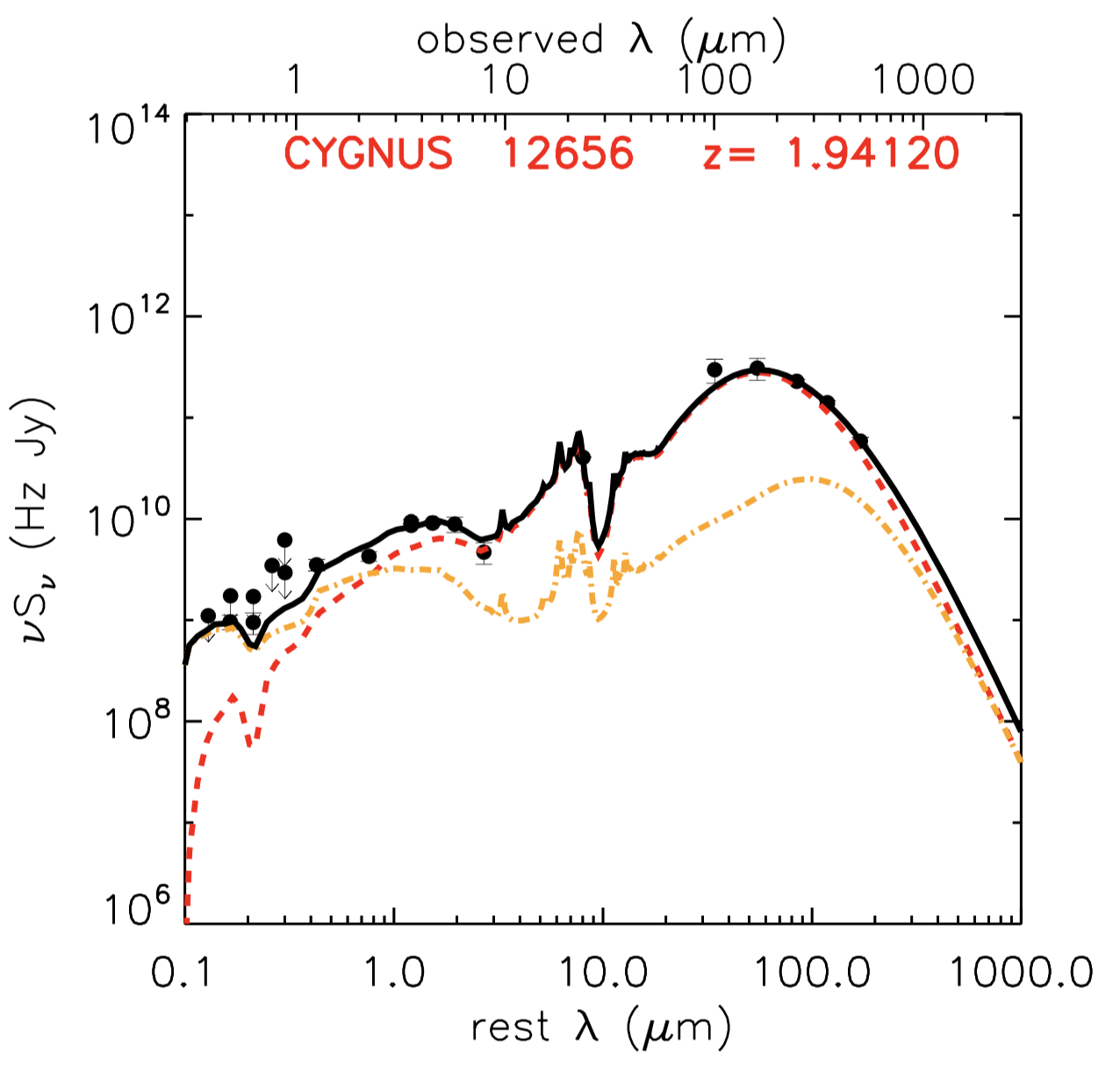}
        \end{subfigure}
        \begin{subfigure}{0.3\textwidth}
                \includegraphics[width=\textwidth]{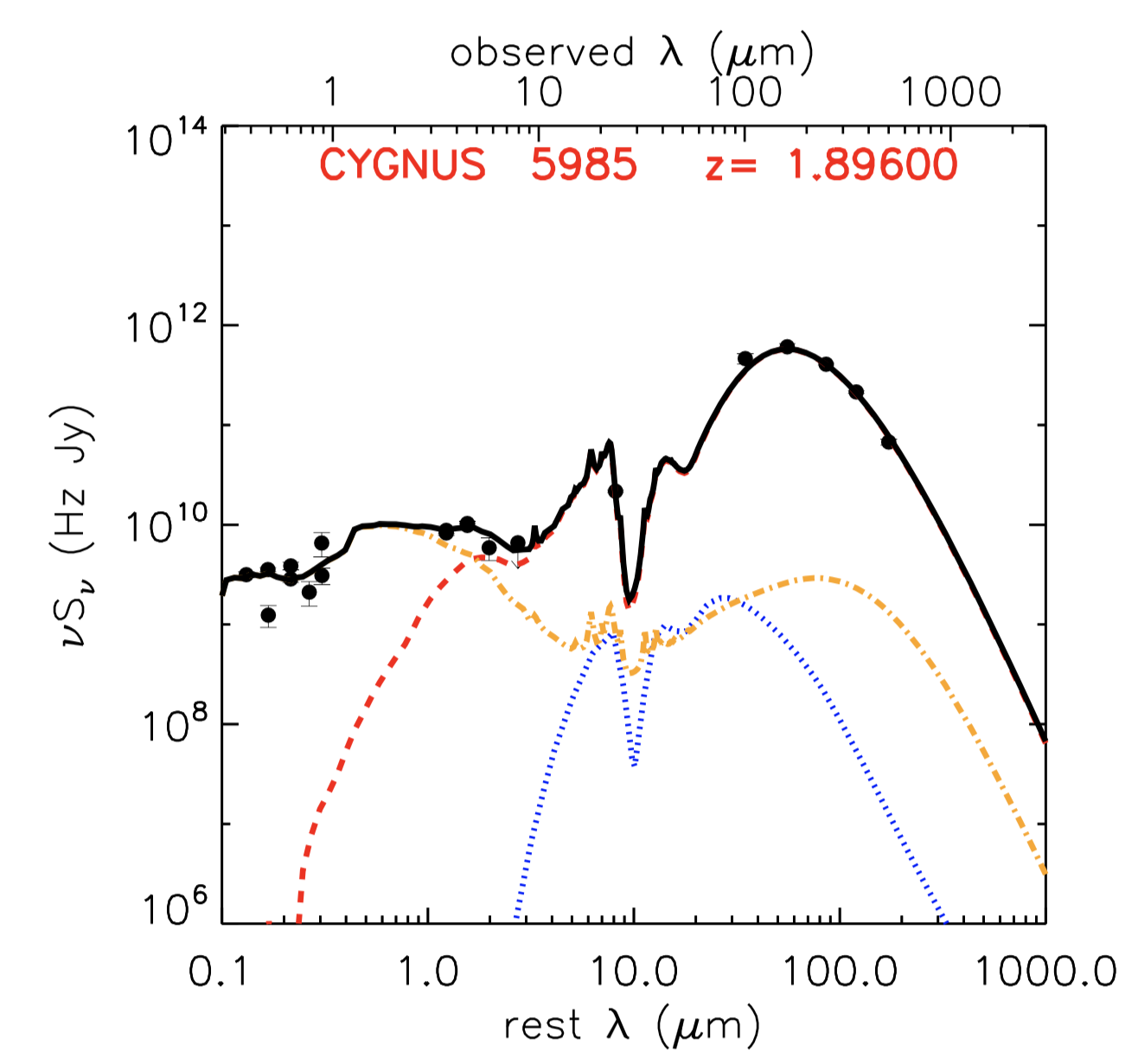}
        \end{subfigure}
        \begin{subfigure}{0.3\textwidth}
                \includegraphics[width=\textwidth]{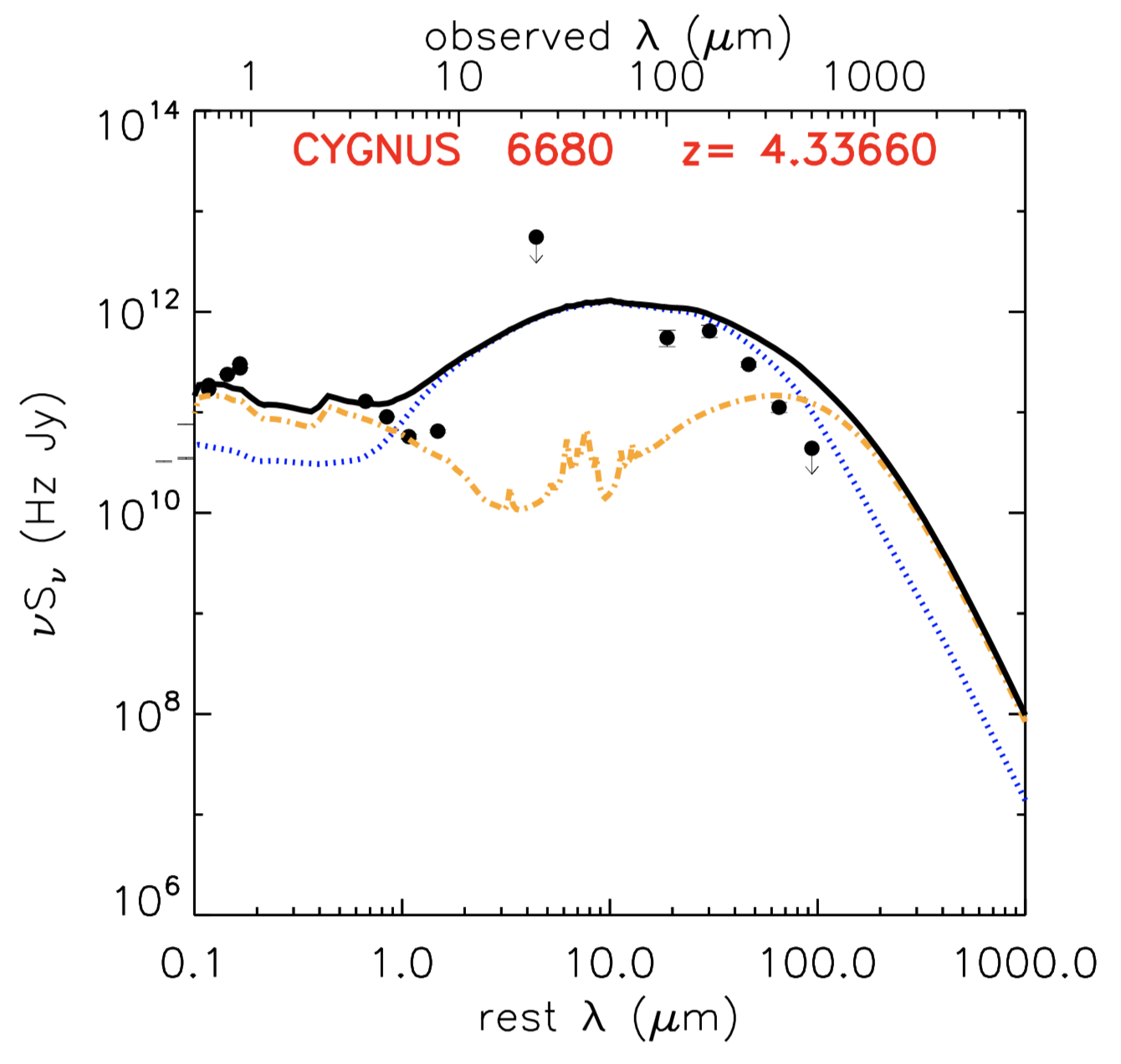}
        \end{subfigure}
\caption{Examples of the best-fit SEDs from CIGALE F06 (first row) and CYGNUS F06 (second row) runs for the same galaxy. First column: Both CIGALE F06 and CYGNUS F06 runs provide good fits. Second column: CIGALE F06 run cannot fit in the FIR band while CYGNUS F06 run gives a good fit. Third column: Both CIGALE F06 and CYNUS F06 runs fail to provide a good fit. Arrows in CYGNUS are flux + 3$\times$error bar when the detection is of low significance (<3 $\sigma$).}
\label{example}
\end{figure*}

Figure \ref{example}  illustrates some examples of the best-fit SEDs from CIGALE F06 and CYGNUS F06 runs for the same galaxy. There are galaxies that both CYGNUS F06 and CIGALE F06 runs can fit well, CYGNUS F06 run provides good fits while CIGALE F06 run gives relatively poor fits and neither can fit well.

\section{SFR on different timescales}\label{timescale}
\begin{figure*}

\resizebox{\hsize}{!}{\includegraphics[width=\linewidth]{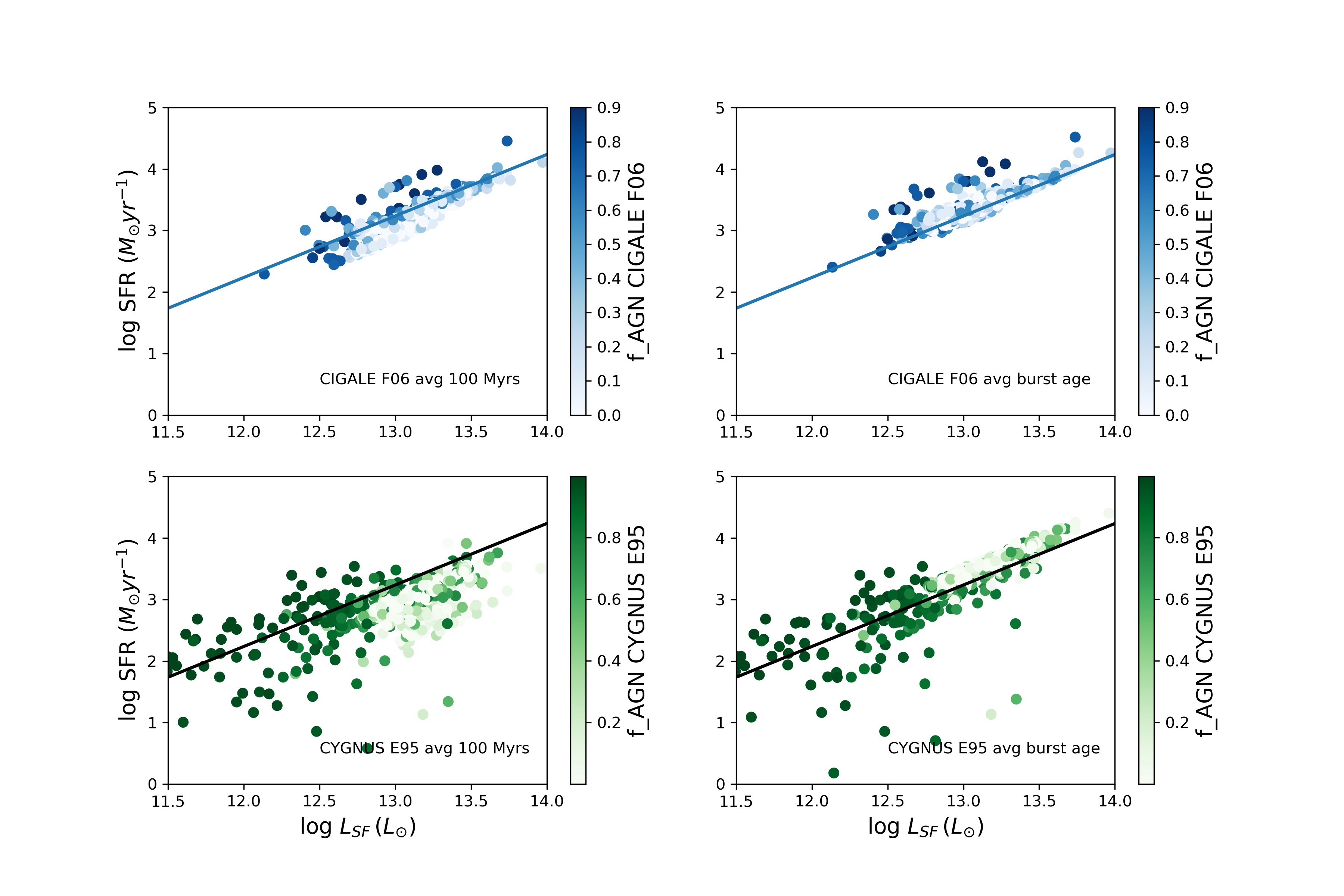}}
\caption{Comparisons of SFR on different timescales for CIGALE F06 (upper panel) and CYGNUS E95 (bottom panel) runs respectively. Left: SFR averaged over 100 mega years compared with IR luminosity due to star formation (denoted as $L_{SF}$), color-coded by the AGN fraction. The solid lines indicate the $L_{SF}$ derived SFR. Right: SFR averaged over burst age compared with $L_{SF}$.}
\label{timescale_comp}
\end{figure*}

In Figure \ref{timescale_comp}, we compare SFR$_{100 Myr}$ and SFR$_{burst\,age}$ to IR luminosity due to star formation ($L_{SF}$) and SFR derived from $L_{SF}$ (denoted as solid lines). We find the correlation between SFR$_{burst\,age}$ and $L_{SF}$ is tighter which is expected since burst age is usually shorter than 100 mega years, leading to less deviation from what the $L_{SF}$ indicates. Some of AGN dominated sources in CIGALE F06 run are more likely to occupy a bit above the solid lines (i.e., higher than $L_{SF}$ derived SFR). In CYGNUS E95 run, SFR$_{100 Myr}$ is systematically lower than SFR derived from $L_{SF}$ while SFR$_{burst\,age}$ is scattered around the solid liness. Using SFR averaged over both timescales will bring contrasting results in Figure \ref{ms_loc} in which high-redshift AGN-dominated sources in CIGALE will mainly reside in an area above the MS, while in CYGNUS they predominantly occupy the area below the MS. This suggests that besides the different methodology adopted, using different SFR indicators may also result in differences among observed trends.

\section{Robustness of the stellar mass estimates}\label{mass_test_t}
\begin{figure*}

\resizebox{\hsize}{!}{\includegraphics[width=\linewidth]{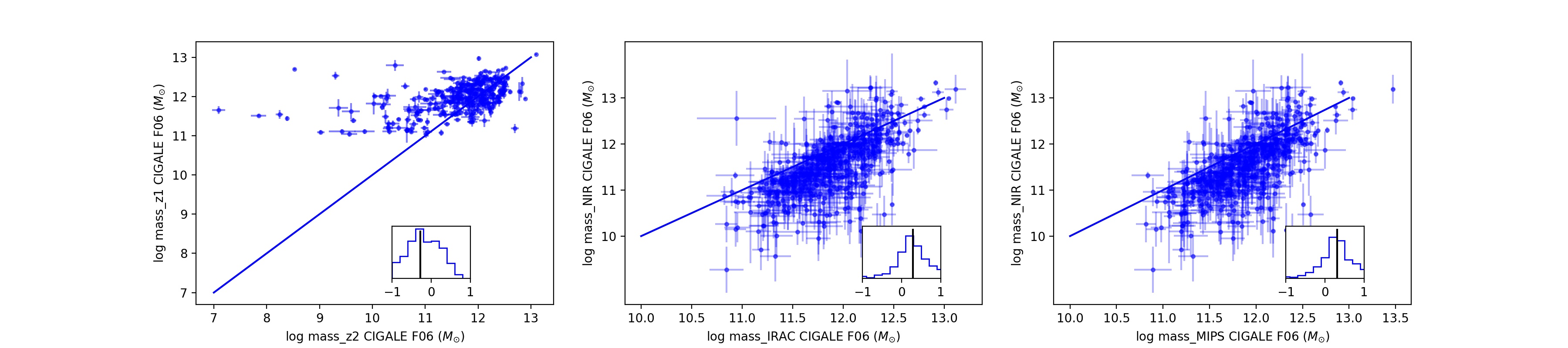}}
\caption{Comparisons of different stellar mass estimates using different redshift or photometric data. Left: Stellar mass estimate comparison using the primary phot-z and secondary phot-z.  Middle: Stellar mass estimate comparison using optical/NIR data only and multi-wavelength photometry up to IRAC. Right: Stellar mass estimate comparison using optical/NIR data only and multi-wavelength photometry up to MIPS.}
\label{mass_test}
\end{figure*}

Another possibility related to some of the ultra-massive galaxies could be due to incorrect photometric redshifts. We select 292 HLIRGs that have a secondary peak above 80\% highest probability density credible interval peak and run CIGALE with F06 model again. They have poorer fitting results compared to primary peak. The left panel of Figure \ref{mass_test} shows the comparison between stellar mass estimates under the primary phot-z and secondary phot-z. 32 \% (68\%) of these HLIRGs have larger (smaller) stellar mass estimates under secondary phot-z compared to stellar mass estimates under primary phot-z respectively. There is a 0.28 dex systematic offset and a 0.33 median absolute deviation (MAD) between the two estimates. There are extreme cases in which the stellar mass estimates using secondary phot-z are two magnitudes above or below the stellar mass estimates using primary phot-z. We can see a long tail towards low mass end in estimates under secondary phot-z. However, there are still 37\% of HLIRGs have stellar mass estimates above $10^{12} M_{\odot}$ under secondary phot-z, compared to 51\% under the primary pho-z. 

Previous studies of global stellar mass function have used multi-wavelength data that are typically shorter than \textit{Spitzer/}MIPS 24 $\mu$m. For example, 
 \citet{2013A&A...556A..55I} included \textit{Spitzer}/IRAC data that only covered $\sim 0.1$ deg$^2$ in the center of COSMOS field, \citet{2014MNRAS.444.2960D} utilized IRAC 3.6 and 4.5 $\mu$m images which only covered rest-frame 0.9 $\mu$m at $z\sim4$. We ran another CIGALE fitting with F06 model using data that only include optical/NIR photometry (excluding bands longer than IRAC) and compare the stellar mass estimates with that derived using multi-wavelength photometry up to IRAC and MIPS 24 $\mu$m respectively. As the middle and right panels of Figure \ref{mass_test} show, the stellar mass estimates using only optical/NIR data are systematically lower than that derived using multi-wavelength photometry up to IRAC and MIPS 24 $\mu$m, with a median value of 0.29 and 0.31 dex respectively. In addition, they have larger errorbars (median value of 0.2 dex compared with 0.09 and 0.08 dex respectively), which means they are relatively less well constrained well.

\begin{figure*}
\centering
\begin{subfigure}{0.3\textwidth}
\includegraphics[width=\textwidth]{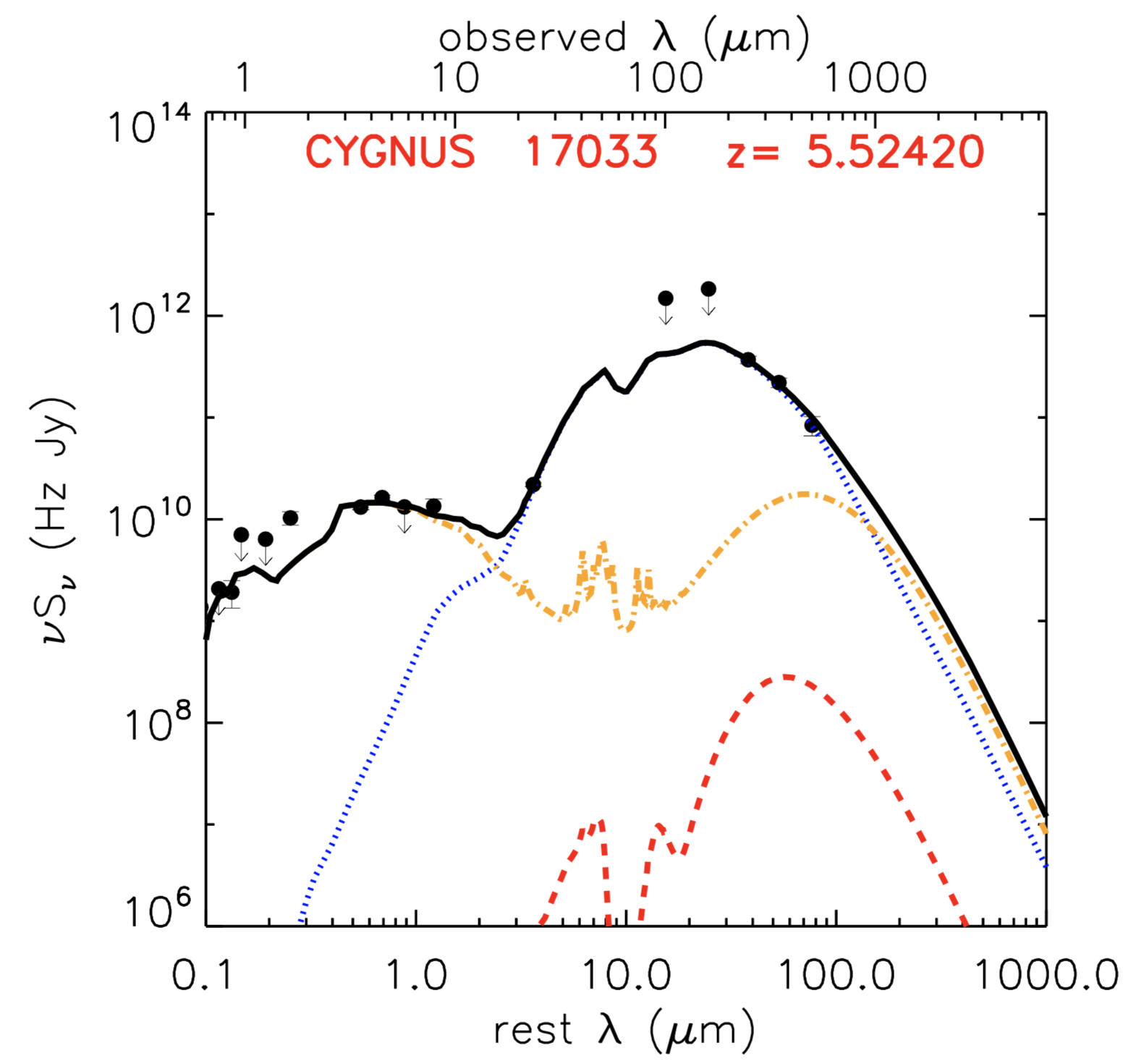}
\end{subfigure}
\begin{subfigure}{0.3\textwidth}
\includegraphics[width=\textwidth]{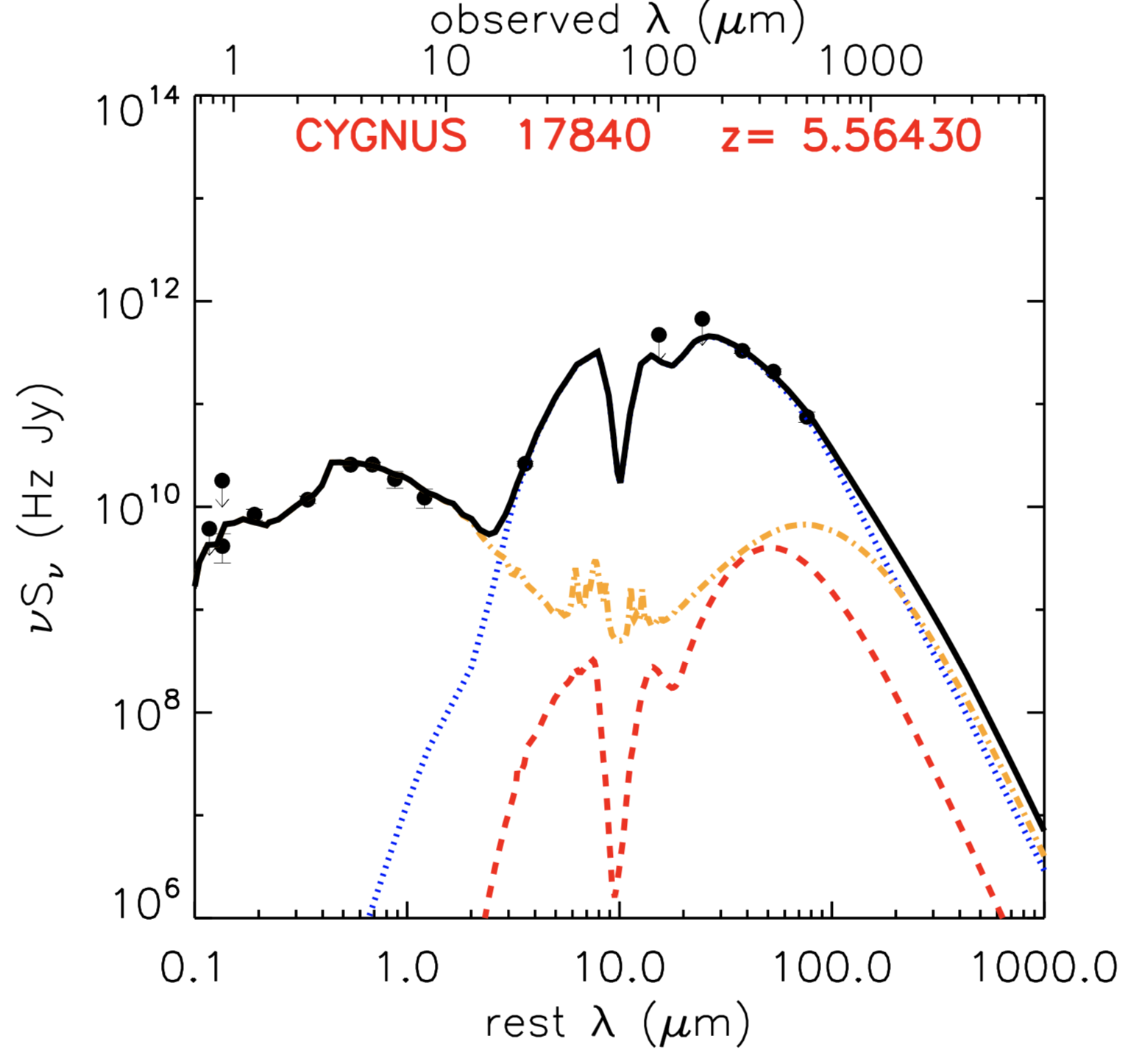}
\end{subfigure}
\begin{subfigure}{0.3\textwidth}
\includegraphics[width=\textwidth]{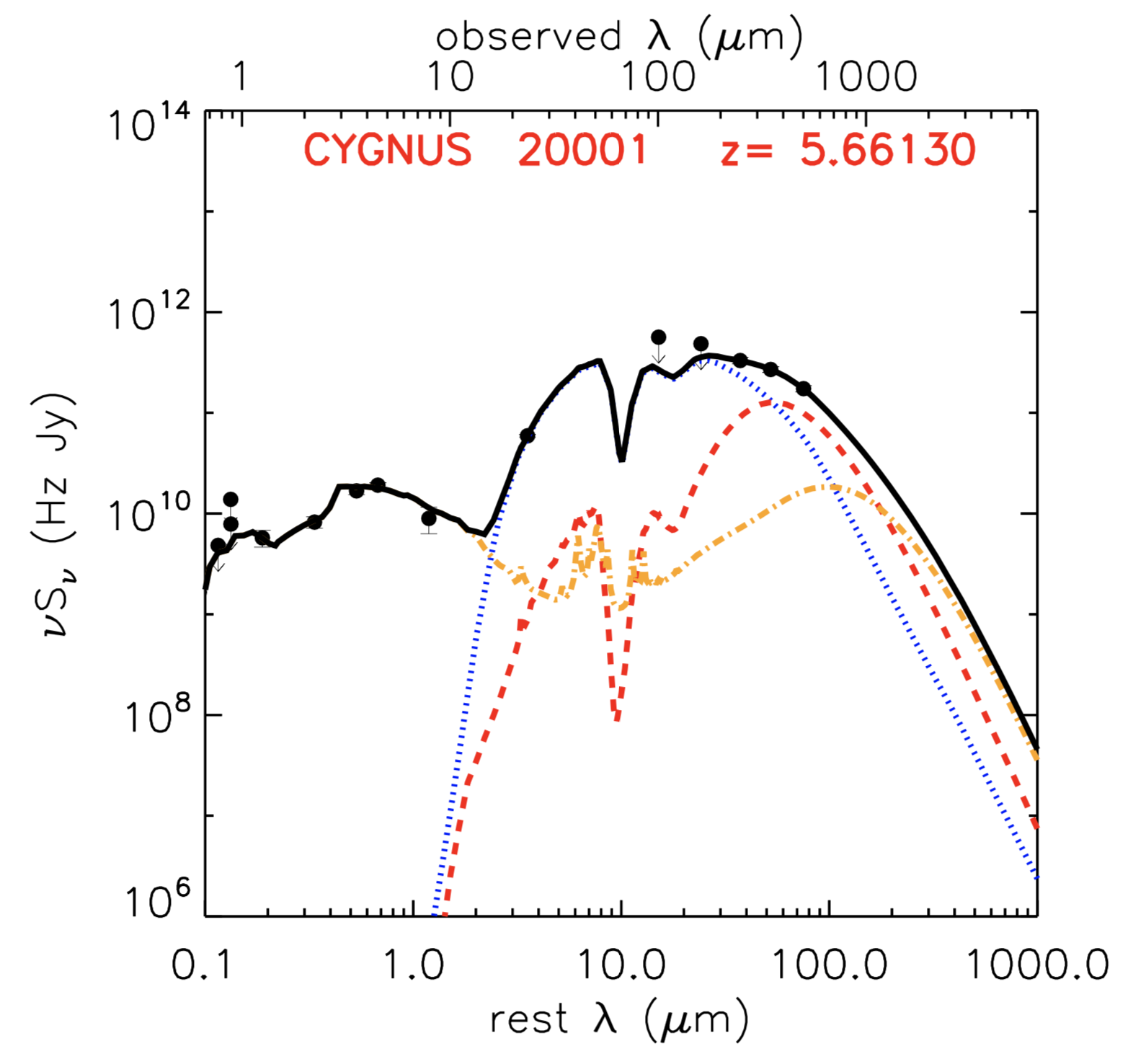}
\end{subfigure}
\caption{Three examples of HLIRGs that have good quality of phot-z and SED fits in CYGNUS E95 run. }
\label{highz}
\end{figure*}

In Figure \ref{highz}, we show three examples of HLIRGs that have good quality of phot-z and SED fits. They are all above $z>5$ and ultra-massive with stellar masses above $10^{12} M_{\odot}$ in all CIGALE and CYGNUS runs.

\section{Examples of HLIRG data products}
In this section, we list six examples (two in each field) of our HLIRGs. In each sample, we include the positions, redshifts, and sources of redshift. The SED fitting results (i.e., stellar mass, SFR, IR luminosity, AGN luminosity, AGN fraction, and the goodness of fit) are following the order of CIGALE F06, CIGALE S12, CYGNUS E95, CYGNUS F06, CYGNUS S12. The full data products will be available online.

\newpage
\begin{sidewaystable}
\caption{Six examples of HLIRGs}
\label{source_list}
\centering
\centering
\begin{tabular}{c c c c c c c c c c c c c }
\hline\hline
id&field&RA (degree)&DEC (degree)&redshift&source&log mass ($M_{\odot}$)&log SFR ($M_{\odot}\, yr^{-1}$)&log $L_{IR}$ ($L_{\odot}$)&log $L_{AGN}$ ($L_{\odot}$)&AGN fraction&reduced $\chi^2$\\
\hline
8030&Bootes&219.2581&34.7748&5.91&phot&12.41$\pm{0.13}$&4.21$\pm{0.03}$&14.10$\pm{0.03}$&13.56$\pm{0.08}$&0.26$\pm{0.05}$&1.61\\
&&&&&&12.45$\pm{0.12}$&4.22$\pm{0.04}$&14.16$\pm{0.04}$&13.69$\pm{0.13}$&0.33$\pm{0.10}$&1.91\\
&&&&&&12.33$\pm{0.23}$&3.91$\pm{0.18}$&14.15$\pm{0.03}$&14.07$\pm{0.10}$&0.96$\pm{0.02}$&1.28\\
&&&&&&12.18$\pm{0.37}$&3.96$\pm{0.17}$&14.09$\pm{0.06}$&13.91$\pm{0.10}$&0.86$\pm{0.07}$&1.09\\
&&&&&&12.17$\pm{0.14}$&4.15$\pm{0.13}$&14.15$\pm{0.03}$&13.88$\pm{0.10}$&0.88$\pm{0.05}$&1.82\\
30041&Bootes&216.6282&33.9191&2.64&spec&11.47$\pm{0.18}$&3.21$\pm{0.12}$&13.08$\pm{0.05}$&13.09$\pm{0.03}$&0.20$\pm{0.00}$&3.77\\
&&&&&&11.50$\pm{0.17}$&3.31$\pm{0.12}$&13.23$\pm{0.03}$&13.35$\pm{0.02}$&0.30$\pm{0.00}$&2.15\\
&&&&&&11.98$\pm{0.11}$&3.24$\pm{0.06}$&13.10$\pm{0.04}$&12.74$\pm{0.13}$&0.36$\pm{0.06}$&6.15\\
&&&&&&10.79$\pm{0.03}$&3.44$\pm{0.00}$&13.24$\pm{0.00}$&13.09$\pm{0.04}$&0.07$\pm{0.01}$&2.69\\
&&&&&&11.87$\pm{0.56}$&3.28$\pm{0.29}$&13.15$\pm{0.27}$&12.95$\pm{0.60}$&0.27$\pm{0.26}$&4.20\\
5155&Lockman&164.9295&58.14691&2.50&phot&11.66$\pm{0.12}$&3.28$\pm{0.08}$&13.16$\pm{0.06}$&12.55$\pm{0.15}$&0.23$\pm{0.10}$&1.69\\
&&&&&&11.74$\pm{0.15}$&3.31$\pm{0.04}$&13.16$\pm{0.04}$&12.42$\pm{0.19}$&0.18$\pm{0.10}$&1.75\\
&&&&&&11.31$\pm{0.30}$&3.39$\pm{0.05}$&13.23$\pm{0.01}$&12.46$\pm{0.10}$&0.31$\pm{0.05}$&1.63\\
&&&&&&11.30$\pm{0.18}$&3.41$\pm{0.04}$&13.21$\pm{0.01}$&12.14$\pm{0.15}$&0.16$\pm{0.19}$&1.84\\
&&&&&&11.99$\pm{0.15}$&3.31$\pm{0.06}$&13.17$\pm{0.01}$&12.56$\pm{0.07}$&0.47$\pm{0.08}$&2.33\\
9658&Lockman&163.9700&57.5173&2.95&spec&12.18$\pm{0.03}$&3.13$\pm{0.02}$&13.27$\pm{0.02}$&13.13$\pm{0.06}$&0.58$\pm{0.05}$&3.68\\
&&&&&&12.20$\pm{0.05}$&3.42$\pm{0.03}$&13.33$\pm{0.02}$&12.83$\pm{0.06}$&0.28$\pm{0.04}$&5.78\\
&&&&&&11.98$\pm{0.11}$&3.31$\pm{0.05}$&13.21$\pm{0.04}$&12.96$\pm{0.09}$&0.26$\pm{0.04}$&2.05\\
&&&&&&10.78$\pm{0.06}$&3.41$\pm{0.01}$&13.27$\pm{0.01}$&13.00$\pm{0.04}$&0.21$\pm{0.02}$&2.81\\
&&&&&&12.07$\pm{0.12}$&3.36$\pm{0.05}$&13.21$\pm{0.02}$&12.88$\pm{0.06}$&0.18$\pm{0.04}$&1.95\\
12488&EN1&245.4060&55.2906&2.68&phot&11.68$\pm{0.09}$&3.31$\pm{0.06}$&13.14$\pm{0.03}$&12.31$\pm{0.15}$&0.13$\pm{0.06}$&1.14\\
&&&&&&11.69$\pm{0.10}$&3.30$\pm{0.03}$&13.10$\pm{0.03}$&12.03$\pm{0.33}$&0.08$\pm{0.09}$&1.39\\
&&&&&&11.81$\pm{0.03}$&3.40$\pm{0.10}$&13.16$\pm{0.04}$&7.97$\pm{4.24}$&0.00$\pm{0.14}$&1.36\\
&&&&&&11.72$\pm{0.12}$&3.37$\pm{0.07}$&13.18$\pm{0.02}$&12.36$\pm{0.20}$&0.13$\pm{0.09}$&1.14\\
&&&&&&11.64$\pm{0.22}$&3.36$\pm{0.13}$&13.15$\pm{0.06}$&12.16$\pm{0.15}$&0.09$\pm{0.13}$&1.14\\
33647&EN1&243.2171&56.2972&1.86&spec&11.49$\pm{0.15}$&3.07$\pm{0.07}$&13.00$\pm{0.03}$&12.67$\pm{0.04}$&0.31$\pm{0.03}$&2.93\\
&&&&&&12.67$\pm{0.15}$&3.18$\pm{0.18}$&13.05$\pm{0.06}$&12.98$\pm{0.02}$&0.21$\pm{0.03}$&3.39\\
&&&&&&11.90$\pm{0.05}$&3.10$\pm{0.02}$&13.02$\pm{0.02}$&12.81$\pm{0.05}$&0.27$\pm{0.02}$&4.20\\
&&&&&&11.78$\pm{0.07}$&3.22$\pm{0.01}$&13.08$\pm{0.02}$&12.93$\pm{0.06}$&0.18$\pm{0.11}$&1.65\\
&&&&&&11.67$\pm{0.08}$&3.11$\pm{0.03}$&13.01$\pm{0.03}$&12.92$\pm{0.09}$&0.33$\pm{0.04}$&2.81\\
\end{tabular}

\end{sidewaystable}

\section{Parameters in SED fitting}
In this section, we list the parameter space in CIGALE (Table \ref{para_CIGALE}) and CYGNUS (Table \ref{para_CYGNUS}), respectively.

\begin{table*}
\caption{Parameter space in CIGALE}
\label{para_CIGALE}      
\centering
\begin{tabular}{c c c}
\hline\hline
model&parameter&values\\
\hline
SFH&e-folding time of the main population&500, 1500, 3500, 5500 Myrs\\
delayed $\tau$+starburst&age the main population&500, 1000,1700, 2300, 3500 Myrs\\
&e-folding time of the late starburst population&10000 Myrs\\
&age of the late starburst population&30,80,120 Myrs\\
&mass fraction of the late starburst population&0.0,0.05, 0.1, 0.2, 0.25, 0.3\\
\hline
SSP&IMF&Salpeter\\
\citet{bc03}&metallicity&solar\\
\hline
dust attenuation&V-band attenuation in the birth clouds&0,  0.1, 0.4, 0.7,  1.1, 1.5, 1.9, 2.5, 3, 3.5\\
double power-law&power-law slope of the attenuation in the birth clouds&-0.7\\
&Av$_{ISM} /$ Av$_{BC}$&0.8\\
&power-law slope of the attenuation in the ISM&-0.7\\
\hline
dust emission&Mass fraction of PAH&1.12\\
\citet{DL14}&minimum radiation field&5,15,40\\
&power-law slope $\alpha$ in dM/dU  $\propto U^\alpha$ &2\\
\hline
AGN templates&Full opening angle of the dust torus&100$\degree$ \\
\citet{Fritz}&viewing angle&10,30,40,60,70,90$\degree$ \\
&AGN fraction&0.0,0.1, 0.2, 0.3, 0.45, 0.6, 0.75, 0.9, 0.99\\
\hline
AGN templates&average edge-on optical depth at 9.7 micron&7, 11\\
SKITOR&half opening angle&40 $\degree$ \\
&viewing angle&10,  30,  40, 60,  70,  90 $\degree$ \\
&AGN fraction&0.0,0.1, 0.2, 0.3, 0.45, 0.6, 0.75, 0.9, 0.99\\
\hline
\end{tabular}
\end{table*} 

\begin{table*}
        \centering
\caption{Parameters of the models used in this paper, symbol used, their assumed ranges and summary of other information about the models. The \citet{Fritz} model assumes two additional parameters that define the density distribution in the radial direction ($\beta$) and azimuthal direction ($\gamma$). In this paper we assume $\beta=0$ and $\gamma=4$. The SKIRTOR model \citep{Stalevski} assumes two additional parameters that define the density distribution in the radial direction ($p$) and azimuthal direction ($q$). In this paper we assume $p=1$ and $q=1$. In addition, the SKIRTOR library we used assumes the fraction of mass inside clumps is 97\%. There are 4 additional scaling parameters for the starburst, spheroidal, AGN and polar dust models, $f_{SB}$,  $f_s$, $f_{AGN}$ and $f_{p}$, respectively. }
        \label{para_CYGNUS}
        \begin{tabular}{llll} 
                \hline
                Parameter &  Symbol & Range &  Comments\\
                \hline
                 &  &  & \\
{\bf CYGNUS Starburst}  &  &  & \\
                 &  &  \\
Initial optical depth of giant molecular clouds & $\tau_V$  &  50-250  &  \citet{1993ApJ...405..538B,bc03}\\
Starburst star formation rate e-folding time       & $\tau_{*}$  & 10-30Myr  &  \cite{E00}, \cite{E09}  \\
Starburst age      & $t_{*}$   &  5-35Myr &  metallicity=solar \\
                   &           &         &    \\
{\bf CYGNUS Spheroidal Host}  &  &  &  \\
                 &  &  \\
Spheroidal star formation rate e-folding time      & $\tau^s$  &  0.125-8Gyr  & \citet{1993ApJ...405..538B,bc03} \\
Starlight intensity      & $\psi^s$ &  1-17 &  \cite{ER03}, \cite{2021MNRAS.503L..11E} \\ 
Optical depth     & $\tau_{v}^s$ & 0.1-15 &  metallicity=40\% of solar\\   
                  &            &  &  \\
{\bf CYGNUS AGN torus}  &  &    &  \\
                 &  &  &  \\
Torus equatorial UV optical depth   & $\tau_{UV}$  &  250-1450 &  Smooth Tapered disks\\  
Torus ratio of outer to inner radius & $r_2/r_1$ &  20-100 & \cite{ER95}, \cite{E13} \\   
Torus half-opening angle  & $\theta_o$  &  30-75\degr & Standard galactic dust mixture \\ 
Torus inclination     & $\theta_i$  &  0-90\degr &  \\ 
                 &            & \\
{\bf Fritz AGN torus}  &  &  &   \\
                 &  &  & \\
Torus equatorial optical depth at 9.7$\mu m$  &  &  0.1-10 & Smooth Flared disks \\  
Torus ratio of outer to inner radius &  &  10-30 &  \cite{Fritz}\\   
Torus half-opening angle  &   &  20-70\degr & Standard galactic dust mixture \\ 
Torus inclination     &   &  0-90\degr &  \\   
                 &            & &  \\
{\bf SKIRTOR AGN torus}  &  &   &  \\
                 &  &  &  \\
Torus equatorial optical depth at 9.7$\mu m$  &   &  3-11 & Two-phase flared disks \\  
Torus ratio of outer to inner radius &  &  10-30 &  \cite{Stalevski}, \cite{2016MNRAS.458.2288S} \\   
Torus half-opening angle  &  &  20-70\degr &  Standard galactic dust mixture\\ 
Torus inclination     &   &  0-90\degr &  \\   
                \hline
        \end{tabular}

\end{table*}

\end{appendix}
\end{document}